\documentclass[prl,twocolumn,showpacs,superscriptaddress]{revtex4-1}

\usepackage{graphicx,bm,amssymb,amsmath}
\usepackage{color}
\usepackage{amsmath}

\begin{document}

\title{Active swarms on a sphere}
\author{Rastko Sknepnek}
\affiliation{Division of Physics and Division of Computational Biology, University
of Dundee, Dundee, DD1 4HN, United Kingdom}
\email{r.sknepnek@dundee.ac.uk}
\author{Silke Henkes}
\affiliation{Institute of Complex Systems and Mathematical Biology, Department of Physics, University of Aberdeen, Aberdeen, AB24
3UE, United Kingdom}
\email{shenkes@abdn.ac.uk}

%\date{\today}

\begin{abstract}
Here we show that coupling to curvature has profound effects on collective motion in active systems, leading to patterns
not observed in flat space. Biological examples of such active motion in curved environments are numerous:
curvature and tissue folding are crucial during gastrulation \cite{vasiev2010modeling}, epithelial 
and endothelial cells move on constantly growing, curved crypts and vili in the gut \cite{ritsma2014intestinal}, and the mammalian corneal epithelium grows in a steady-state vortex pattern \cite{collinson_clonal_2002}. On the physics side, droplets coated with actively driven microtubule bundles show active nematic 
patterns \cite{sanchez2012spontaneous}. We study a model of self-propelled particles with polar alignment on a sphere. Hallmarks of these motion patterns are a \emph{polar vortex} and a \emph{circulating band} arising due to the incompatibility
between spherical topology and uniform motion - a consequence of the ``hairy ball'' theorem. We present analytical results showing that
frustration due to curvature leads to stable elastic distortions storing energy in the band.
\end{abstract}

\maketitle

Active systems have recently attracted a flurry of interest \cite{vicsek2012collective,marchetti2013hydrodynamics}.
Each particle is equipped with its own source of energy that enables motility. The system is characterized by a constant 
input of energy at the individual particle level, rendering it out of equilibrium. The local energy input, many-body effects 
and dissipation result in a variety of motion patterns. Examples span multiple length scales ranging from the microscale, \emph{e.g.}, bacterial colonies \cite{sokolov2007concentration}, migration of tissue cells \cite{szabo2006phase}
and motion of the cytoskeleton \cite{juelicher2007active} to the macroscales, \emph{e.g.}, fish schools \cite{hemelrijk2005density}, bird flocks \cite{bajec2009organized}, 
migrating mammals \cite{fischhoff2007social}. Important examples on the non-living side include active nematic fluids \cite{giomi2013defect,thampi2014instabilities},
active colloidal swimmers \cite{palacci_living_2013}, vibrating granular disks \cite{deseigne2010collective} and traffic \cite{helbing2001traffic}.

Being far from equilibrium limits the statistical mechanics description of active systems. Instead, one resorts either
to hydrodynamic models \cite{marchetti2013hydrodynamics} or to simulations \cite{vicsek2012collective}. 
A lot of insight was gained by studying toy systems beginning with Vicsek \emph{et al.} \cite{vicsek1995novel}, who constructed a model of constant velocity self-propelled particles (SPP)
that noisily align with their neighbours. Soon after, a hydrodynamic description was constructed using symmetry arguments \cite{toner1995long} and later derived microscopically \cite{bertin2006boltzmann}.
A silent point in the Vicsek model is that particles are point-like and align instantaneously. The model can be extended to include
excluded volume, but its effects remain poorly understood, especially at high 
densities \cite{henkes2011active,Tailleur2008,Fily2012,Bialke2012,berthier2013non}.
It is, however, known that models with volume exclusion can form stable vortex states in two- and three-dimensional flat space \cite{DOrsogna2006,Strefler2008}.
Geometry can play a profound role in many systems. A prominent examples is the structure of the ground states of crystals
on curved surfaces \cite{baush2003grain}. Curvature effects are not only limited to static properties, but are also expected 
to affect the dynamics. It is intuitively clear that it is not possible to have a uniform-velocity fluid flow of a sphere, and a similar argument 
applies to active systems in curved geometries: A flock on a sphere cannot take a conformation with all particles travelling at the same speed.

\begin{figure}
\begin{centering}
\includegraphics[width=0.95\columnwidth]{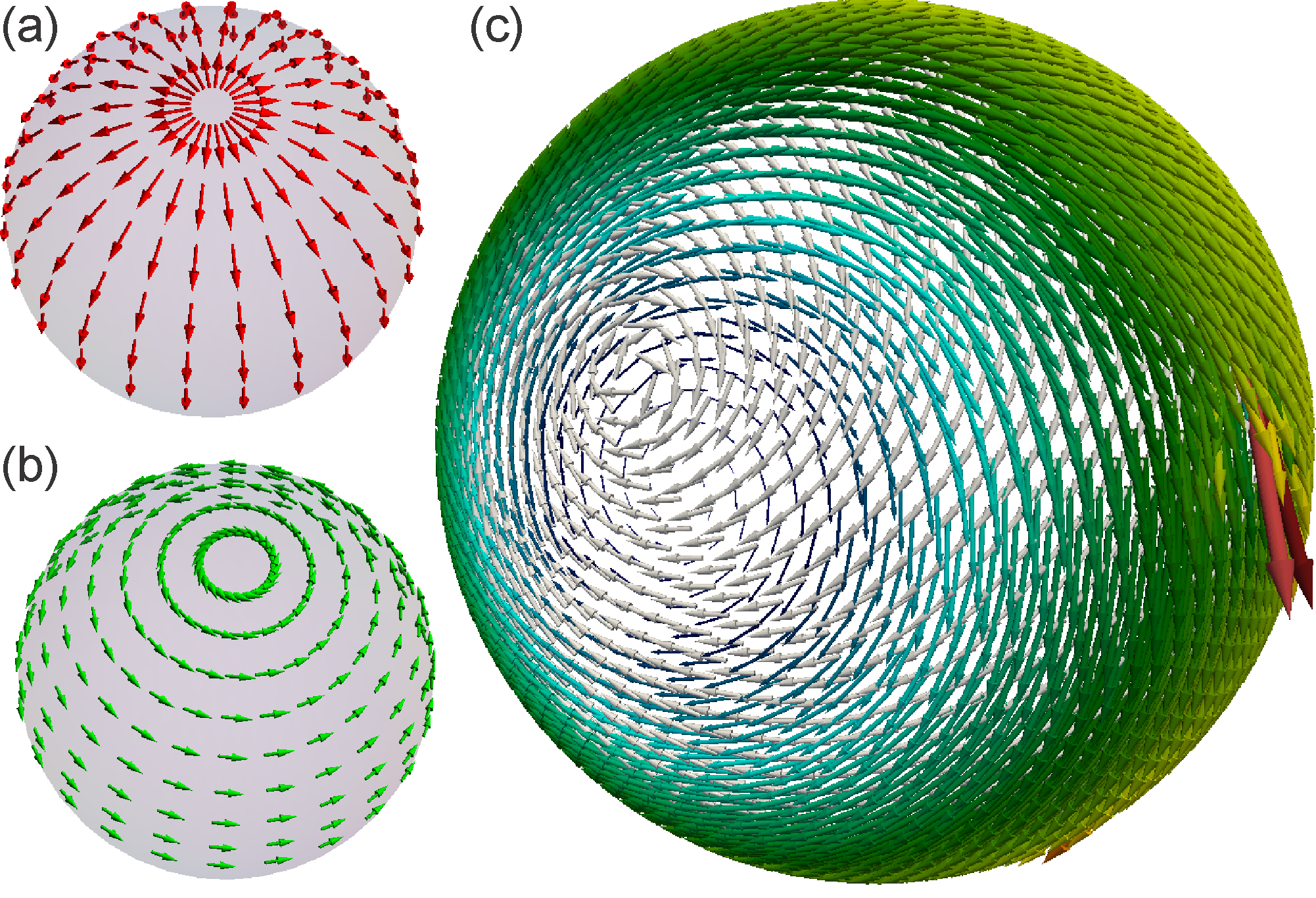}
\end{centering}
\caption{Two possible vector field configurations around a $+1$ topological defect on a sphere: {\bf a.}~source and {\bf b.}~whirlpool vortex. {\bf c.} Motion pattern of 
active particles on a sphere. Coloured arrows indicate velocity vectors, $\mathbf{v}_i$, with the colour proportional to $\left|\mathbf{v}_i\right|$.
Gray arrows represent particles' direction vectors, $\mathbf{n}_i$. For low activity $\mathbf{v}_i$ and $\mathbf{n}_i$ are not necessarily aligned. Note that only the whirlpool, {\bf b}~is
consistent with polar active motion on the sphere.}
\label{fig:vortex_drawing}
\end{figure}

All SPP models to date have assumed a flat geometry. In this letter we examine self-propelled particles confined to move on a sphere subject 
to a realistic alignment rule and white noise. We draw inspiration from recent experiments of Sanchez, \emph{et al.} \cite{sanchez2012spontaneous}. 
Our goal here is not to describe those experiments, which requires consideration of hydrodynamic effects, but to construct a minimal model, which provides clear insight into the interplay between activity and geometry.

\begin{figure}
\begin{centering}
\includegraphics[width=0.99\columnwidth]{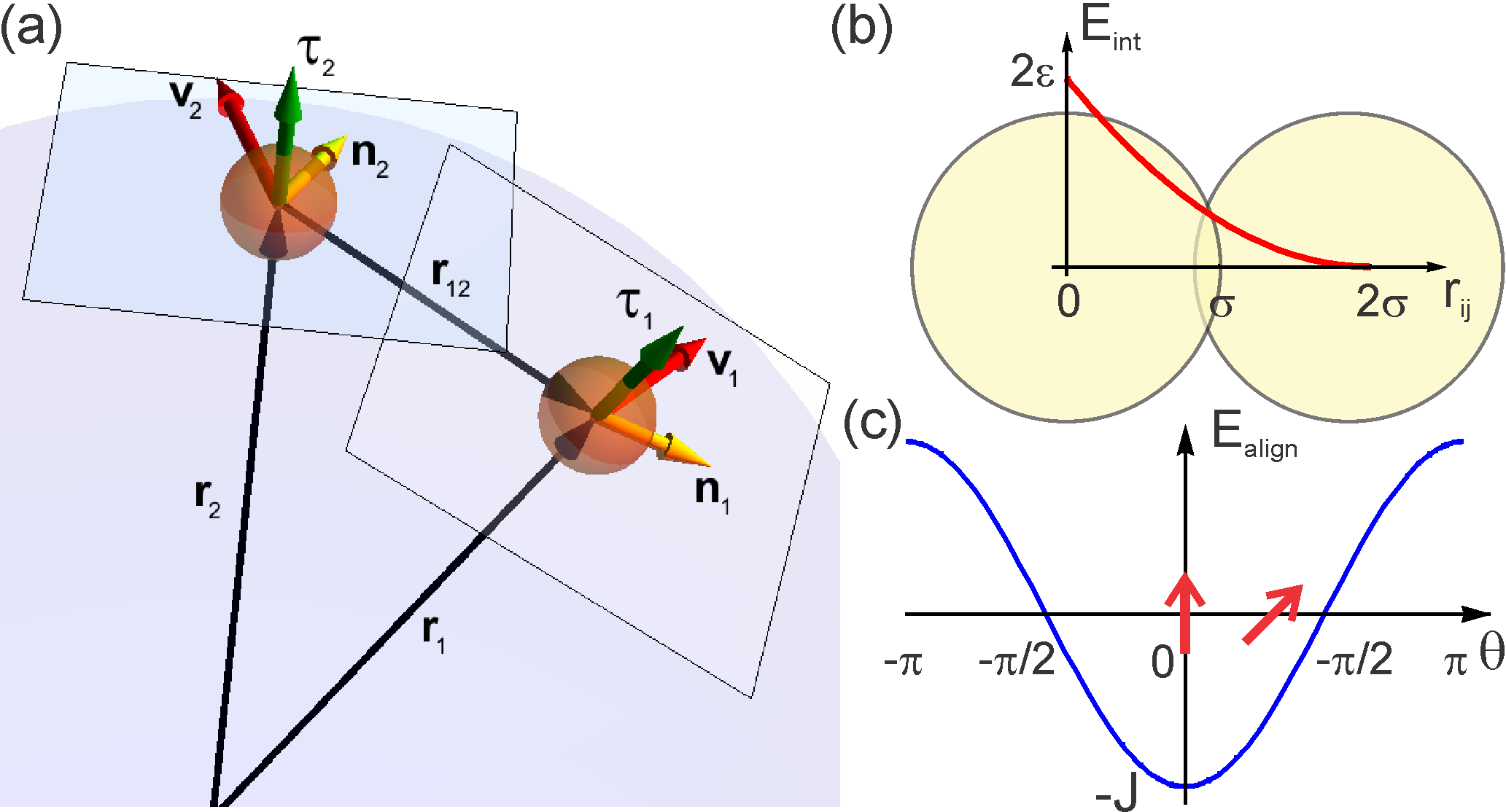}
\par\end{centering}

\caption{{\bf a.} Schematic representation of the system. Particles are modelled as spheres of radius $\sigma$ confined to move on the surface of a sphere of radius $R$.
Particles' centres are described by radius vectors $\mathbf{r}_{i}$ and each particle is endowed with a unit-length direction vector $\mathbf{n}_{i}$,
which can point in an arbitrary direction but is confined to the tangent plane at $\mathbf{r}_{i}$. The velocity vector $\mathbf{v}_{i}$ is in
general not parallel to its direction, but is also confined to the tangent plane; then the torque $\boldsymbol{\tau}_{i}$ exerted on
each particle points along the normal vector at $\mathbf{r}_{i}$. The Euclidean distance $r_{ij}$ between particles is computed in the embedding
$\mathbb{R}^{3}$ space.  {\bf b.} Particles interact via a short-range soft potential, which is finite for any value of $r_{ij}$. {\bf c.} Particle alignment is assumed to follow
the XY model with ferromagnetic coupling constant $J$. \label{fig:model}}
\end{figure}

Our system consists of $N$ spherical particles of radius $\sigma$ confined to the surface of a sphere of radius $R$ (Fig.~\ref{fig:model}{\bf a}).
Particle velocity, $\mathbf{v}_i$, and direction, $\mathbf{n}_i$, are constrained to the tangent plane at every point. In the overdamped limit, the equations of motion are (see SI)
\begin{eqnarray}
\dot{\mathbf{r}}_{i} & = & \mathbf{P}_T\!\left(\!\mathbf{r}_i, v_{0}\mathbf{n}_{i}+\mu\sum\nolimits_{j}\mathbf{F}_{ij}\!\!\right)\label{eq:projection_r}\\
\dot{\mathbf{n}}_{i} & = & \left[\text{P}_N\!\left(\!\mathbf{r}_i,-J\sum\nolimits_{j}\mathbf{n}_{i}\times\mathbf{n}_{j}\!\right)+\xi_{i}\right]\left(\hat{\mathbf{r}}\times\mathbf{n}_i\right),\label{eq:projection_n}
\end{eqnarray}
where $v_{0}$ is the self-propulsion velocity pointing along $\mathbf{n}_{i}$. The interaction force $\mathbf{F}_{ij}$ is modelled as a short-range repulsion, $\mathbf{F}_{ij}=-k\left(2\sigma-r_{ij}\right)\frac{\mathbf{r}_{i}-\mathbf{r}_{j}}{r_{ij}}$ 
for $r_{ij}<2\sigma$ and $\mathbf{F}_{ij}=0$ otherwise, with $k$ being the elastic constant. $r_{ij}$ is the Euclidean distance computed in $\mathbb{R}^{3}$ and $\mu$ is mobility. 
 Alignment follows XY-model dynamics with coupling constant $J>0$ and the sum is carried over all neighbours within a $2.4\sigma$ cutoff radius, \emph{i.e.} the first shell of neighbours.  
 $\mathbf{P}_T\left(\mathbf{r}_i, \mathbf{a}\right) = \mathbf{a}-\left(\hat{\mathbf{r}}_i\cdot\mathbf{a}\right)\hat{\mathbf{r}}_i$ and 
$\text{P}_N\left(\mathbf{r}_i,\mathbf{a}\right) = \left(\hat{\mathbf{r}}_i\cdot\mathbf{a}\right)$ are, respectively,  projection operators of vector $\mathbf{a}$ onto the tangent plane and the normal vector at $\mathbf{r}_i$.
Particle orientation is subject to delta-correlated noise $\xi_{i}$ acting in the tangent plane with strength
$\nu_{r}$. An important feature of our model is the separate dynamics of $\mathbf{n}_{i}$ and $\mathbf{v}_{i}$ \cite{szabo2006phase}. In the absence of
interactions, $\mathbf{n}_{i}$ and $\mathbf{v}_{i}$ will eventually align. The interparticle forces, however, 
allow for permanent deviations of $\mathbf{v}_{i}$ from $\mathbf{n}_{i}$, a key mode for active elastic energy storage \cite{henkes2011active}.
The coupling constant $J$ sets an alignment time scale, $\tau_{al}\approx1/J$. Similarly, the collision time scale is set by
$k$ as $\tau_{col}\approx1/\mu k\Delta$, where $\Delta$ is the maximum overlap with respect to $\sigma$. In the following,
length is measured in units of $\sigma$, energy in units of $k\sigma^2$, time in units of $\tau=1/\mu k$, velocity in units 
of $\sigma/\tau\equiv\mu k \sigma$, and $\nu_{r}$ in units of $\tau^{-1}$. Finally, equations~(\ref{eq:projection_r}) and (\ref{eq:projection_n}) are integrated numerically (see \emph{Materials and Methods}).

\begin{figure*}
\begin{centering}
\includegraphics[angle=90,width=0.85\textwidth]{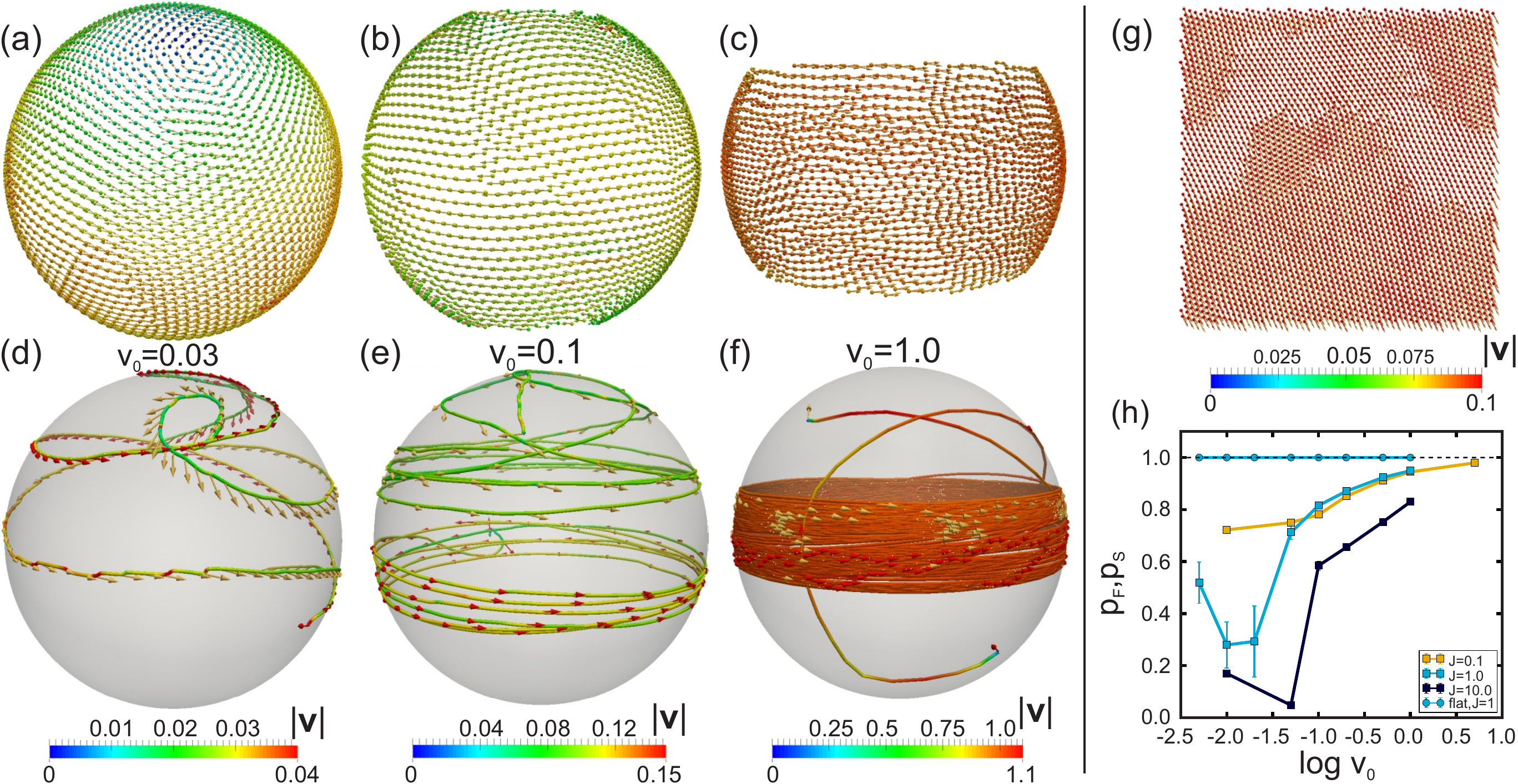}
\par\end{centering}
\caption{Steady state configurations at $t=10^{4}\tau$ for {\bf a.}~$v_{0}=0.03\sigma/\tau$, {\bf b.}~$v_{0}=0.1\sigma/\tau$ and {\bf c.}~$v_{0}=1\sigma/\tau$, with $J=1\tau^{-1}$ and $\nu=0.002\tau^{-1}$, see also SI movies.
The length and colour of velocity vectors reflect the magnitude of $\left|\mathbf{v}_{i}\right|$. Thinner yellow vectors indicate the directions of the orientation
vectors $\mathbf{n}_{i}$. For clarity, only particles on the front hemisphere are shown and the relative length of the velocity vectors between panels is not preserved.
Panels {\bf d.}, {\bf e.}~and {\bf f.}~show trajectories of two randomly selected particles coloured according to $\left|\mathbf{v}_i\right|$. Vectors along the trajectory indicate the direction of the orientation vector
at selected subsets of positions. {\bf g.}~is a snapshot of the $v_{0}=0.1\sigma/\tau$ periodic flat system of size $L=100\sigma$; here $\mathbf{v}_i$ (red) is uniform and completely aligned with $\mathbf{n}_i$ (yellow).
Panel {\bf h.}~shows the order parameters for the flat ($p_F$) and spherical ($p_S$) systems as a function of $v_0$ for a range of values of $J$. 
\label{fig:traj}
}
\end{figure*}

Fig.~\ref{fig:traj} shows snapshots of typical motion patterns for $v_0=0.03\sigma/\tau$, $0.1\sigma/\tau$ and $1.0\sigma/\tau$. We focus
on the low noise ($\nu=0.002\tau^{-1}$) and large packing fraction ($\phi=1$) regime. For low $v_0$ one observes a polar vortex pattern (Fig.~\ref{fig:traj}{\bf a}). In this state, spherical symmetry is spontaneously broken and two vortices form at opposite poles (see Fig.~\ref{fig:vortex_drawing}{\bf b}). The entire flock rotates around the axis passing through those poles.
This circulating band has neither sources nor sinks, as required for a particle conserving fluid, so only the pattern in Fig.~\ref{fig:vortex_drawing}{\bf b} is permitted. Linear velocity within the flock is not uniform, gradually decreasing from $v_0$ at the equator to zero towards the poles. In general, $\mathbf{n}_i$ is not aligned with $\mathbf{v}_i$ and forms separate vortices
(grey arrows in Fig.~\ref{fig:vortex_drawing}{\bf c}). The motion is heavily frustrated with short-lived localized velocity spikes and rearrangements (longer arrows in Fig.~\ref{fig:vortex_drawing}{\bf c}) leading to substantial mixing as can be
 seen in individual particle trajectories (Fig.~\ref{fig:traj}{\bf d}). As $v_0$ increases, the system develops ``bald'' spots at the poles. Particles 
are compressed towards the equator and the flock takes the configuration of a spherical belt. 
%The velocity within the band is not uniform, decreasing away from the centre. 
$\mathbf{n}_i$ and $\mathbf{v}_i$ are more closely aligned and there are fewer jumps in velocity. Finally, as $v_0$ is increased to
$1.0\sigma/\tau$, the flock is squeezed further towards the equator. The velocity distribution within the flock is nearly uniform and $\mathbf{n}_i$ and $\mathbf{v}_i$ are almost aligned.
Particle trajectories are very regular (Fig.~\ref{fig:traj}{\bf f}).

% A flat-space counterpart of our system exhibits both a polar and a disordered phase \cite{szabo2006phase}, but has not been fully characterized.
Local reductions of velocity due to volume exclusion and decoupling of $\mathbf{n}_i$ and $\mathbf{v}_i$ lead to active phase separation \cite{Tailleur2008,Fily2012}, an effect distinct from the banding observed here:
We have examined the flat-space counterpart of our system in the same range of values of $v_0$ and $J$ as in the spherical case. It remains in the homogenous phase (Fig.~\ref{fig:traj}{\bf g} and SI movie). 
Using a Vicsek order parameter $p_F = \frac{1}{N v_0} \left| \sum_{i}  \mathbf{v}_i\right|$, we show that this flat system is also consistently in the polar phase, with $p_F\approx 1$ independent of $v_0$ (Fig.~\ref{fig:traj}{\bf h}). In the spherical case now, we measure alignment on the surface of the sphere. We define 
$p_S = \frac{1}{N R v_0} \left| \sum_{i} \mathbf{r}_i \times \mathbf{v}_i \right|$, $p_S\rightarrow 1$ for a circulating ring moving at $v_0$. $p_S$ transitions from a low value for small $v_0$ to near perfect alignment at larger $v_0$ (Fig.~\ref{fig:traj}{\bf h}). 
This shows that the transition to the polar vortex and a moving band is a purely curvature-driven effect, with no equivalent in the planar model.

The phenomenon is similar to the ring structures found in the plane \cite{DOrsogna2006} and in three dimensions \cite{Strefler2008}, with the important difference that in here it occurs in the absence of attraction. Active contractile elements have also been studied in a continuum model on a cylinder and show banding \cite{srivastava_patterning_2013}. We note that without the self avoidance (\emph{i.e.},~at $k=0$), our model reduces to a continuum Vicsek model. In contrast
to the polar ordered state observed on the plane, on the sphere after a long relaxation period the entire flock collapses into a ring spanning one of the great circles. The effect again differs from the density banding close to the Vicsek transition \cite{gregoire_onset_2004} since it occurs deep inside the polar regime.

\begin{figure*}[t]
 \centering
 \includegraphics[angle=270,width=0.85\textwidth]{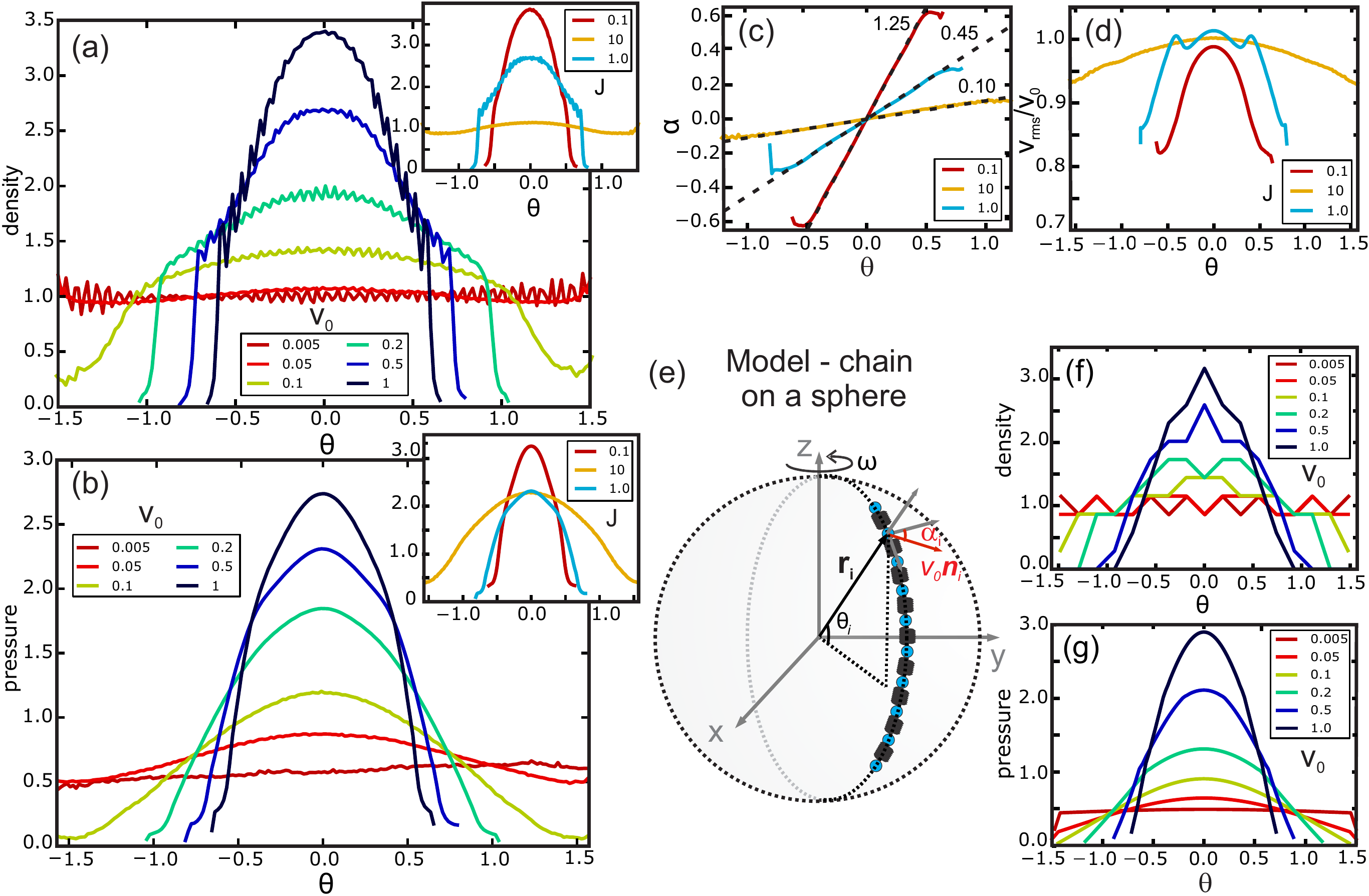}
\caption{{\bf a.} Density profiles for increasing $v_0$ at $J=1\tau^{-1}$. \emph{Inset}: density profiles as a function of $J$, for $v_0=0.5\sigma/\tau$. {\bf b.} Pressure profiles (virial part), same parameters as in {\bf a}.
 {\bf c.} Angle $\alpha$ of the self-propulsion direction with the equator, as a function of $J$ for $v_0=0.5\sigma/\tau$. Dashed lines are linear fits, with slopes denoted by the values.
 {\bf d.} Root-mean-square velocity profiles, same parameters as {\bf c} and {\bf e}. Sketch of the chain on a sphere model and local coordinate system, see text.  {\bf f.} and {\bf g.} Density and pressure profiles
 (virial part) for $J=1\tau^{-1}$ predicted using energy minimization of the model with the fitting parameter $\text{s}=0.55$ for $J=1\tau^{-1}$.}
 \label{fig:profiles}
\end{figure*}

We focus only on the high velocity regime with a developed band.
In Figs.~\ref{fig:profiles}{\bf a} and \ref{fig:profiles}{\bf b}, we present the density and pressure profiles in the established band for $J=1\tau^{-1}$ and a range of $v_0$. The density has been normalized 
to that of a uniformly covered sphere, and we measure pressure by computing the trace of the local \emph{force moment tensor}, $\hat{\Sigma}_i = \sum_j \mathbf{r}_{ij} \mathbf{F}_{ij}$ with units of energy (see SI).
The band has a relatively complex structure.
For example, the discrete particles lead to a distinct layering pattern in the density profiles. Similarly, a change of slope in the pressure profiles stems from double overlaps among very compressed particles,
though, overall, the band narrows and the pressure increases with growing $v_0$. The influence of $J$ is at first counterintuitive 
(Fig.~\ref{fig:profiles}, insets): the \emph{lower} values of $J$ where the alignment is weaker lead to more pronounced banding! 

To understand this, we analyse the active mechanics of an already formed band. We consider a slice cut out of the sphere in the polar direction (Fig.~\ref{fig:profiles}{\bf e} and SI). 
The particles in the slice all move in the same tangential direction, with decreasing speed towards the poles. In local spherical coordinates the particle position is 
$\mathbf{r}_i = R(\cos \theta_i \cos \phi_i, \cos \theta_i \sin \phi_i, \sin \theta_i)$, where $z$ is the polar direction,
 $\theta_i$ is the angle from the $xy$-plane along a meridian and $\phi_i$ is the azimuthal angle. 
Since the active force $\mathbf{F}_i^{\text{act}}=v_0 \mathbf{n}_i$ is always in the tangential plane, we can write
$\mathbf{n}_i = \cos \alpha_i \mathbf{e}_{\phi} - \sin \alpha_i \mathbf{e}_{\theta}$. Here $\alpha_i$ is the (signed) angle between the polar direction and the band velocity along the equator.
We derive a simple relation between rotation speed and active forces by projecting equation~(\ref{eq:projection_r}) onto the $\phi$ direction. Assuming steady state rotation with angular velocity $\omega$, 
we have $\dot{\mathbf{r}}_i=R \omega \cos\theta_i \mathbf{e}_{\phi}$, leading to (see SI):
\begin{equation}
\cos \alpha_i  = \frac{R \omega}{v_0} \cos \theta_i. \label{eq:alpha_theta} 
\end{equation}
This corresponds to a symmetric vector field pointing inwards to both sides of the equator (visible in Fig.~\ref{fig:traj}{\bf c}).
In Fig.~\ref{fig:profiles}{\bf c},~we show simulation results for $\alpha$ vs.~$\theta$ profiles, for three different values of the alignment parameter $J$. All profiles are linear, 
with a slope that depends only on $J$ (see SI). Since $\alpha_i$ is also the angle between the polar direction and the velocity, it now makes sense that $\alpha$ \emph{reduces} for 
large values of $J$. With $s$ the slope of the graph, we have $\alpha=s\theta$, with $s\approx 1.25, 0.45$ and $0.1$ for $J=0.1\tau^{-1},1\tau^{-1}$
 and $10\tau^{-1}$, respectively. In Fig.~\ref{fig:profiles}{\bf d}, we show the velocity magnitude profiles for the same runs. In all cases, 
velocities reach near or above $v_0$ at the centre of 
the band and then reduce towards the edges, but are more complex than the simple parabolic profile predicted by equation~(\ref{eq:alpha_theta}).

Along the chain, in the direction $\mathbf{e}_{\theta}$, we can find an approximate form of strain $u_s$ using a force-on-a-chain method (see SI).
To leading order, the strain is given by 
\begin{equation} 
u_s(\theta) = -\frac{v_0}{\sigma \mu k}\left[\frac{\cos(s \theta)-\cos (s\theta_m)}{\kappa s} + \sin(s\theta_m)\right],
\end{equation}
which we use to extract density and pressure profiles. Here $\theta_m$ is the location of the band edge, itself a model output (equation~(28) of SI). Assuming a homogeneous system (see SI), the pressure (virial part) is given by the stress-strain
relation $p=\bar{k} u_s$ ($\bar{k}$ is an effective stiffness), and density $\rho/\rho_0 \approx 1 - u_s$, where $\rho_0$ is the initial density.
 $\kappa = 2\sigma/R$ is the dimensionless curvature of the sphere. Negative strain indicaties increased density and inward pressure, consistent with a compressed band. 
The inward pressure at the edges, $p = -\frac{v_0}{\sigma \mu} \sin(s\theta_m)$ is equal and opposite to the active force per unit length, $\frac{v_0}{\sigma \mu} \sin \alpha$ due to the self propulsion, 
that is pressure balance reminiscent of active phase separation \cite{Fily2012,Tailleur2008} and a first order phase transition. From our analysis, four important dimensionless parameters emerge: 
the reverse alignment strength $s$, the underlying curvature $\kappa\approx0.07$, the active pressure $v_0/\sigma \mu$ and the density through $\theta_m$ (see SI). To achieve the quantitative fit of the pressure and density profiles in Fig.~\ref{fig:profiles}{\bf f} and {\bf g}, we use a discrete energy minimization approach (see \emph{Materials and Methods}).

In this letter we have constructed and analysed a simple model for overdamped polar active particles confined to move on the surface of a sphere and subject to volume exclusion and a realistic alignment rule.
Using numerical simulations and analytical arguments we have shown that activity and curvature combine to produce interesting types of active patterns: a polar vortex and a stable rotating band structure, not present in the flat case.
While the current approach omits hydrodynamic interactions that may play a role in some experimental systems, it provides a valuable insight into
the intricate, yet poorly understood interplay between curvature and dynamics far from equilibrium. In this study we focused on a narrow range of parameters and yet found a rich set of 
motion patterns, purely driven by geometry. We hope that our results will motivate further
experimental and theoretical studies in this direction in order to shine more light onto this highly biologically relevant problem.   

\section*{Materials and Methods}

Equations of motion (equations~(\ref{eq:projection_r}) and (\ref{eq:projection_n})) were integrated numerically. Instead of choosing a curvilinear
 parametrization of the sphere we kept the equations in the vector form and imposed constraints after each step. Each time step has two stages:
\emph{i}) unconstrained move and \emph{ii}) projection onto the constraint. First, the particle is moved according to equation~(\ref{eq:projection_r}) without
any constraints. Its position is then projected back onto the sphere and its velocity and orientation are projected
onto the tangent plane at the new position. Similarly, torques were projected onto the surface normal at $\mathbf{r}_{i}$ and, finally, $\mathbf{n}_{i}$
was rotated by a random angle around the same normal. As long as the time step is sufficiently small, all projections are unique and
should not affect the dynamics.  

The packing fraction, $\phi=N\pi\sigma^{2}/4\pi R^{2}$ is defined as the ratio of the area occupied by all particles to the total area of the sphere (we count double overlaps twice). All simulations were performed with $N\approx3\times10^{3}$
particles at packing fraction $\phi=1$, resulting in $R\approx28.2\sigma$. For comparison, we performed a series of simulations in the plane with
the same $N$ and $\phi$ by imposing periodic boundary conditions onto a square simulation box of size $L=100\sigma$. In all cases, the equations
of motion were integrated for a total of $1.1\times10^{4}\tau$ with time step $\delta t=10^{-3}\tau$. Initially, particles were placed
at random on the sphere. In order to make the configuration reasonably uniform and avoid large forces leading to large displacements, initial
overlaps were removed by using a simple energy relaxation scheme (with $v_0=0$) for $10^{3}\tau$ time steps. 
Subsequently, activity and noise were introduced and equations were integrated for addition $10^{4}\tau$ using a standard Euler-Maruyama
method. Configurations were recorded every $5\tau$. Typical runs took approximately 5 hours on a single core of Intel Xeon E2600 series processor.

The system spontaneously breaks spherical symmetry and there is no reason to expect that the axis connecting poles will be aligned with any
of the coordinate axes in $\mathbb{R}^3$. Therefore, in order to produce the angular profiles in Fig.~\ref{fig:profiles}, for each snapshot we first
 determined the direction of the total angular velocity and then performed a global rotation around the origin that aligned it with the
$z$-axis in $\mathbb{R}^3$.

In order to analyse the single-slice model we suppose that the chain consists of $N_p$ particles pole-to-pole. We chose $N_p$ such that $p\sigma^2\approx 0.5 k$ in the absence of activity, consistent with the low velocity and flat value of the pressure (see SI). Assuming overlapping particles, the force an adjacent particle $j$ exerts on particle
$i$ in the chain is given by $\mathbf{F}_{ij}=-k\hat{\mathbf{r}}_{ij}(2\sigma-|\mathbf{r}_j-\mathbf{r}_i|)$. $k$ is the (linearised) stiffness of the potential and $\sigma$ is the particle radius. 
If we introduce curvilinear coordinates along the chain and expand around $\theta_i$ in small values of $\delta \theta =\theta_j - \theta_i$, we can approximate
$\mathbf{r}_j-\mathbf{r}_i=-R(\theta_j-\theta_i)\hat{\mathbf{e}}_{\theta}$. To first order, interparticle forces are along $\hat{\mathbf{e}}_{\theta}$, and the forces
acting on particle $i$ from its neighbours $i-1$ and $i+1$ are $F_{i,i-1}=k(2\sigma-R(\theta_i -\theta_{i-1}))$ and $F_{i,i+1}=-k(2\sigma-R(\theta_{i+1}-\theta_i))$. Finally, we can
write the set of equations of motion along the chain:
\begin{align}
 & v_0 \sin \alpha_1 = -\mu k \left(2\sigma-R(\theta_2-\theta_1)\right) \nonumber \\
 & v_0\sin \alpha_i = -\mu k R(\theta_i -\theta_{i-1}) +\mu kR(\theta_{i+1}-\theta_i) \nonumber \\
 & v_0\sin \alpha_{N_p} =  \mu k\left(2\sigma-R(\theta_{N_p} -\theta_{N_p-1})\right).  
\label{eq:band_balance}
\end{align}
We solve these equations using two approaches. First, we treat equations~(\ref{eq:band_balance}) as Euler-Lagrange equations of an energy functional containing only potential energy terms, which we then minimize by using
the standard L-BFGS-B conjugate gradient method including boundary constraints. Formally, even though our physical system conserves neither energy nor momentum, if we assume $\alpha = s \theta$, the active force components in equation~(\ref{eq:band_balance})
derive from an effective potential $V^i_{\text{act}}=v_0 \cos(s \theta_i)$ which can be added to the interparticle repulsive term $V^i_{\text{rep}}=\frac{k R}{2} \sum_{j \in \mathcal{N}} (\theta_j - \theta_i)^2$. 
Then setting the gradients of $V^i = V^i_{\text{act}}+V^i_{\text{rep}}$ to zero is equivalent to equations~(\ref{eq:band_balance}). 
The second approach is based on the analytical continuum limit. It is less straightforward, but a bit more insightful and discussed in details in the SI. 
\acknowledgments{{\bf Acknowledments.} We thank M.C. Marchetti for introducing us to active matter, and for illuminating discussions and critical reading of the manuscript. We also thank F.~Ginelli for useful discussions. Part of this work was performed at the Kavli Institute for Theoretical Physics and was supported in part by the National Science Foundation under Grant No.~NSF PHY11-25915.}

% \end{document}
% \clearpage
\newpage

% \onecolumn
% \noindent
% {\large Supplementary information for \emph{Active swarms on a sphere}}
% \twocoumn
% \vspace*{1cm}
% \begin{widetext}
%  {\large Supplementary information for \emph{Active swarms on a sphere}}
% \end{widetext}
\onecolumngrid
\begin{center}
 {\large SUPPLEMENTARY INFORMATION}
 \end{center}
 \vspace*{2cm}
\twocolumngrid

\section{Constraint motion on a sphere: Holonomic constraints}

This first section derives the correct equations for active, self-propelled constraint motion on a
sphere. The basis for the treatment below can be found, \emph{e.g.}~in Leimkuehler
and Reich  \cite{leimkuhler_simulating_2004}. Consider the following Newtonian full equations
of motion in three dimensions for the spatial variables, $\mathbf{r}_i$: 
\begin{equation}
m\ddot{\mathbf{r}}_{i}=-\gamma\dot{\mathbf{r}}_{i}+\sum_{j}\mathbf{F}_{ij}+\mathbf{F}_{i}^{\text{act}}.\label{eq:fullmotion}
\end{equation}
Here the active force $\mathbf{F}_{i}^{\text{act}}$ is treated as
an independent parameter.

In standard Hamiltonian dynamics, a holonomic constraint is a constraint
which does not depend on the generalized velocities $\dot{q}_{i}$ and can be expressed
as a function of the generalized coordinates $q_{i}$ only. If such a constraint
$\alpha$ is written as as an equation $g_{\alpha}(\mathbf{q})=\mathbf{0}$ ($\mathbf{q}=\left\{q_1,\dots,q_N\right\}$, where $N$ is the total number of degrees of freedom),
$g_{\alpha}(\mathbf{q})$ can be interpreted as a potential, and the
constraint trajectories will then lie on the isopotential surface
with potential value $0$. The spherical constraint $g(\mathbf{r})=x^{2}+y^{2}+z^{2}-R^{2}$
(with $R$ being the radius) is a classic example of such a constraint.

Using a reasoning similar to electrostatics or gravitation, the constraint
forces keeping the system on its isopotential surface need to be normal
to this surface. In other words, they must be along the gradient of
$g$, so that for each constraint, there exists a constraint force
$\mathbf{F}_{\alpha}=\lambda_{\alpha}\nabla_{\mathbf{q}}g_{\alpha}(\mathbf{q})$
that penalizes any deviations from the isopotential surface.

Then for a set of constraints $\lbrace g_{\alpha}(\mathbf{q})=0|\alpha=1,\dots,M\rbrace$,
and an explicitly Hamiltonian system, the equations of motion are
 \cite{leimkuhler_simulating_2004}: 
\begin{align}
\frac{d\mathbf{q}}{dt} & =\mathbf{v}\nonumber \\
m\frac{d\mathbf{v}}{dt} & =-\nabla_{\mathbf{q}}V(\mathbf{q})-\sum_{\alpha}\lambda_{\alpha}\nabla_{\mathbf{q}}g_{\alpha}(\mathbf{q}).\label{eq:constraint}
\end{align}
To determine the multipliers $\lambda_{\alpha}$, we can take further
derivatives of the constraint equations: 
\begin{equation}
\frac{d}{dt}(g_{\alpha}(\mathbf{q}))=\nabla_{\mathbf{q}}g_{\alpha}(\mathbf{q})\cdot\mathbf{v}=0.
\end{equation}
As to be expected, this shows that $\mathbf{v}$ belongs to the tangent
bundle of the constraint surface $g_{\alpha}(\mathbf{q})$. Finally,
to determine $\lambda_{\alpha}$, we can differentiate this equation
once more, and then substitute the equations of motion, equation (\ref{eq:constraint}).
We should then obtain a set of $M$ equations to determine the $M$
multipliers $\lambda_{\alpha}$. Depending on our choice of constraints,
these equations will be linearly independent, and offer an unique
set of $\lambda_{\alpha}$. 

Even though the active part of equation (\ref{eq:fullmotion}) does
not derive from a potential, the steps outlined above remain valid. We choose the set of positions $  \lbrace \mathbf{r}_i \rbrace \equiv \mathbf{q}$ as generalized coordinates.
The gradient of our constraint $g(\mathbf{r}_i)=x_i^{2}+y_i^{2}+z_i^{2}-R^{2}$
is $\nabla_{\mathbf{r}_i}g(\mathbf{r}_i)=2\mathbf{r}_i$. Then the constraint
equations of motion become 
\begin{equation}
m\ddot{\mathbf{r}}_{i}=-\gamma\dot{\mathbf{r}}_{i}+\sum_{j}\mathbf{F}_{ij}+\mathbf{F}_{i}^{\text{act}}-2\lambda_i\mathbf{r}_{i}.\label{eq:constraint_sphere}
\end{equation}
Note that the constraint applies to each particle independently and, thus, $\lambda$ has index $i$.
The derivative constraint just leads to the equation $\dot{\mathbf{r}}_{i}\cdot\mathbf{r}_{i}=0$.
If we define the unit normal to the sphere as $\hat{\mathbf{r}}_i=\mathbf{r}_i/|\mathbf{r}_i|=\mathbf{r}_i/R$,
this confirms that the velocity has to be tangential to the surface of
the sphere.

The second derivative constraint finally allows us to determine $\lambda_i$,
and after substituting equation (\ref{eq:constraint_sphere}) we obtain:
\begin{equation}
2\lambda_i=\frac{1}{r_{i}^{2}}\left[mv_{i}^{2}+\mathbf{r}_{i}\cdot(\mathbf{F}_{i}^{\text{act}}+\sum_{j}\mathbf{F}_{ij})\right].
\end{equation}
Then, after substituting $\lambda_i$ back into equation (\ref{eq:constraint_sphere}),
we can finally write equations of motion that fully implement the
spherical constraint: 
\begin{align}
m\ddot{\mathbf{r}}_{i}=&-\gamma\dot{\mathbf{r}}_{i}+\sum_{j}\mathbf{F}_{ij}+\mathbf{F}_{i}^{\text{act}} \\
&-\frac{\mathbf{r}_{i}}{r_{i}^{2}}\left[m\dot{r}_{i}^{2}+\mathbf{r}_{i}\cdot(\mathbf{F}_{i}^{\text{act}}+\sum_{j}\mathbf{F}_{ij})\right].\nonumber
\end{align}

In the overdamped limit, we can see that $m\rightarrow0$ does not
produce any singularities and we can write the valid equations of motion:
\begin{equation}
\gamma\dot{\mathbf{r}}_{i}=\mathbf{F}_{i}^{\text{act}}-(\hat{\mathbf{r}}_{i}\cdot \mathbf{F}_{i}^{\text{act}})\mathbf{r}_{i}+\sum_{j}\mathbf{F}_{ij}-(\hat{\mathbf{r}}_{i}\cdot \mathbf{F}_{ij})\mathbf{r}_{i}.
\end{equation}
This is in the end a very simple equation, which projects active and
passive forces onto the sphere. If we define the projection operator
at a point $\mathbf{r}_i$ on the sphere acting on a vector $\mathbf{a}$
as $\mathbf{P}_{T}(\mathbf{r}_i,\mathbf{a})=\mathbf{a}-(\hat{\mathbf{r}}_i\cdot\mathbf{a})\hat{\mathbf{r}}_i$,
the overdamped equations of motion are simply 
\begin{equation}
\gamma\dot{\mathbf{r}}_{i}=\mathbf{P}_{T}(\mathbf{r}_{i},\mathbf{F}_{i}^{\text{act}}+\sum_{j}\mathbf{F}_{ij}).\label{eq:projection}
\end{equation}

\section{Angular dynamics}
Fundamentally, we would like to implement a two dimensional $XY$-model type dynamics, where a particle aligns explicitly with its neighbours. In two dimensions, using first order dynamics, we have
\begin{equation} \dot{\phi}_i = -J \sum_j \sin(\phi_i - \phi_j)+\xi_i, \label{eq:angular_xy0} \end{equation}
where $\phi$ is the angle of $ \mathbf{n}$ with the $x$-axis, \emph{i.e.}~$ \mathbf{n}_i=(\cos \phi_i,\sin \phi_i)$ and the first term on the RHS is simply the torque. We have also added a scalar delta-correlated angular noise with distribution $\langle \xi_i(t)\xi_j(t')\rangle = \sigma^2 \delta_{ij}\delta(t-t')$.
On the sphere, it is not possible to define $\phi$ globally and uniquely for each tangent plane, so we need to write the equation in terms of $ \mathbf{n}_i$ directly. The RHS of equation (\ref{eq:angular_xy0})
can be written as a curl projected along the $\mathbf{e}_z$ axis orthogonal to the $xy$-plane to obtain its magnitude:
\begin{equation} 
\dot{\phi}_i = -J \left(\sum_j  \mathbf{n}_i\times  \mathbf{n}_j\right)\cdot \mathbf{e}_z+\xi_i. \label{eq:angular_xy1} 
\end{equation}
On the sphere now, if we define the normal projection of a vector on the unit normal to the tangent plane as $\text{P}_N(\hat{\mathbf{r}_i},\mathbf{a})=(\mathbf{a} \cdot \hat{\mathbf{r}}_i)$, 
the deterministic part of the left hand side of equation (\ref{eq:angular_xy1}) is simply $\text{P}_N(\hat{\mathbf{r}}_i, -J \sum_j  \mathbf{n}_i\times  \mathbf{n}_j)$. The derivative of a unit vector is an angular rotation, and we have $\frac{d \mathbf{n}_i}{dt}=\dot{\phi}_i (\hat{\mathbf{r}}_i \times  \mathbf{n}_i) $, that is the time derivative is both orthogonal to the axis of rotation and the vector itself. Then the $XY$-like angular dynamics on the sphere is given by:
\begin{equation} 
\frac{d \mathbf{n}_i}{dt}=\left[\text{P}_N(\hat{\mathbf{r}}_i, -J \sum_j  \mathbf{n}_i\times  \mathbf{n}_j)+\xi_i\right] (\hat{\mathbf{r}}_i \times  \mathbf{n}_i).
\end{equation}

We note that the fully vectorial approach is as well beneficial from the point of view of numerical simulations as it is straightforward to generalize to an arbitrary surface, unlike working with
local parametrizations, which often have singular points (\emph{e.g.}~for $\theta=0$) and can be costly to compute numerically.

\section{Steady-state rotating solution}

In the simulations, we observe a steady-state rotating solution where
particles cluster symmetrically around the equator and the
whole flock performs a solid-like rotation with angular velocity $\omega$
around an axis through the poles.

\begin{figure}[h]
\centering{}\includegraphics[trim = 10mm 105mm 10mm 0mm,clip,width=0.75\columnwidth]{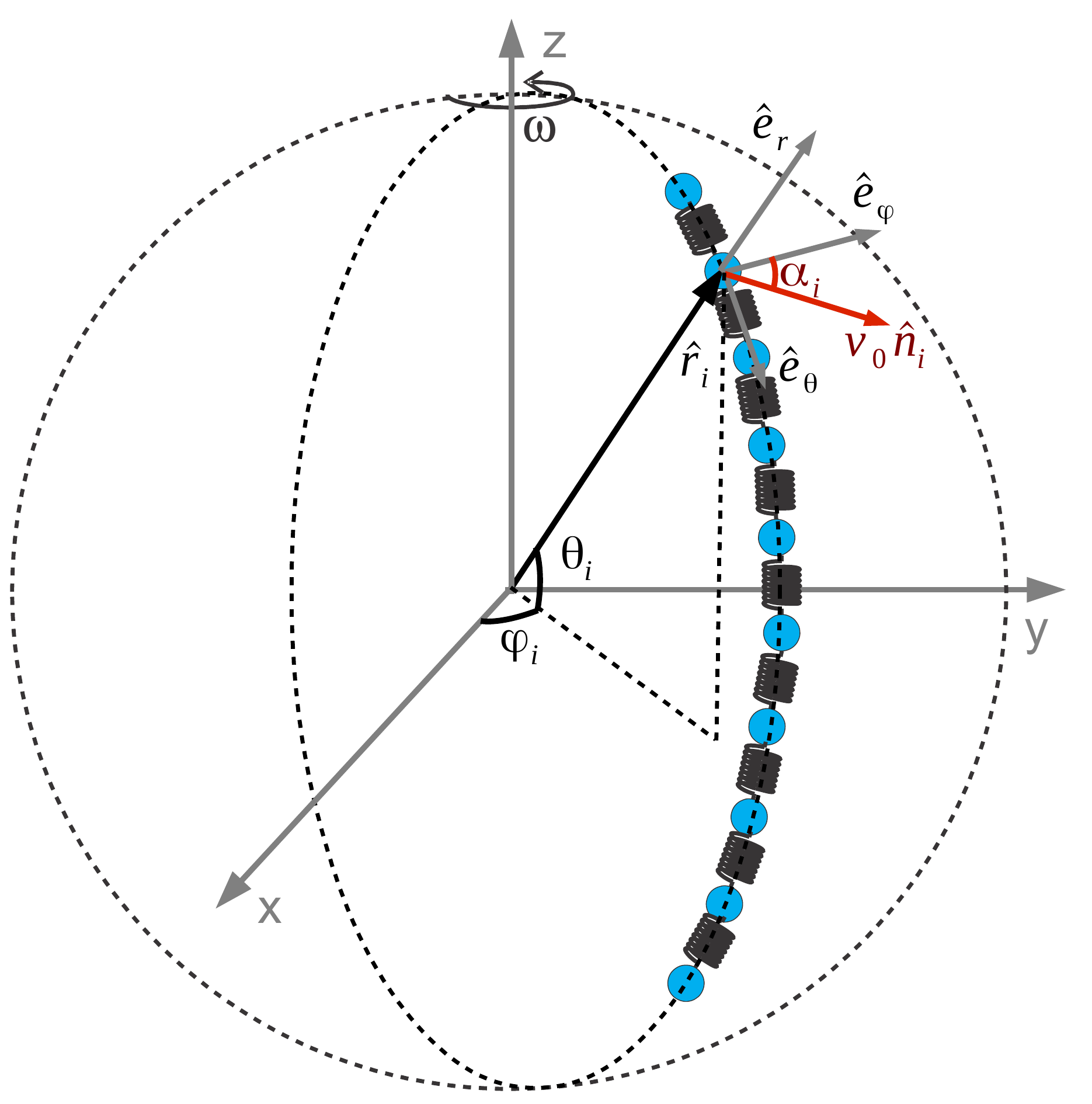}
\protect\caption{Linked spring chain model for a slice of the rotating solution. \label{fig:chain}}
\end{figure}

Before we proceed, let's first set notations and dimensions in order. In the same
units and notation as in the main paper, we have: 
\begin{equation}
\dot{\mathbf{r}}_{i}=\mathbf{P}_T(\mathbf{r}_{i},v_{0} \mathbf{n}_{i}+\mu\sum_{j}\mathbf{F}_{ij}),\label{eq:projection2}
\end{equation}
where we have set $\mathbf{F}_i^{\text{act}} = v_0\gamma \mathbf{n}_{i}$. $v_{0}$ is the constant magnitude of the self-propulsion velocity pointing along $\mathbf{n}_{i}$, and we have mobility $\mu=1/\gamma$.

To approach this situation, consider a one-particle wide ``orange''
slice cut out of the sphere in the polar direction, as shown in Figure
\ref{fig:chain}. All particles in that slice move in the same tangential
direction, with decreasing speed towards the poles. While they are
effectively constrained to a great circle, we unfortunately \emph{cannot}
use the machinery of holonomic constraints derived above since the
constraint condition depends on the velocities $\dot{q}_{i}$,
not only the positions. The additional spherical constraint is of
course still described by the projection in equation (\ref{eq:projection}).

If we move to spherical coordinates, the position of a single particle
is given by \\
$\mathbf{r}_{i}=R(\cos\theta_i\cos\phi_i,\cos\theta_i\sin\phi_i,\sin\theta_i)$,
where $z$ is the polar direction, $\theta_i$ is the angle from
the $xy$-plane and $\phi$ is the azimuthal angle measured from an
arbitrary $x$ axis. The unusual choice of $\theta_i$ is such that
$\theta=0$ corresponds to the equator, and the band reaches between
$-\theta_{m}$ and $\theta_{m}$. We can choose $\phi_i=0$ without
loss of generality, leading to $\mathbf{r}_{i}=R(\cos\theta_i,0,\sin\theta_i)$.

Let us first analyse the active forces. Since equation (\ref{eq:projection})
shows that the projected $\mathbf{F}_{i}^{\text{act}}=v_{0}\gamma \mathbf{n}_{i}$
has to be in the tangential plane, we can write $ \mathbf{n}_i=\cos\alpha_i \mathbf{e}_{\phi}-\sin\alpha_i \mathbf{e}_{\theta}$,
where the first part is in the direction of rotation, and the second
one is along the great circle ( $ \mathbf{e}_{\theta}=[\sin\theta_i,0,-\cos\theta_i]$).
Here $\alpha_i$ is the angle between the polar direction and the particle
velocity.

\begin{figure}[h]
\centering
\includegraphics[clip,width=0.99\columnwidth]{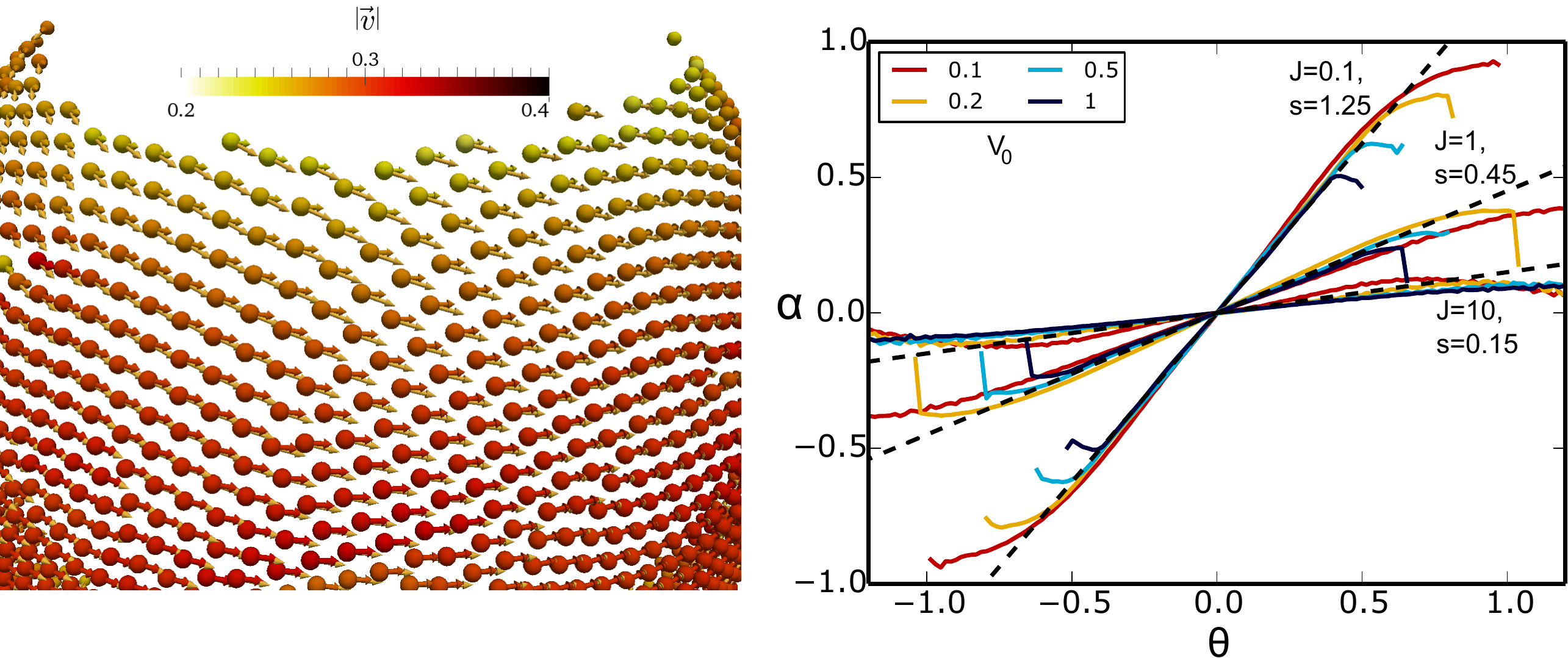}
\protect\caption{Left: Rotating steady state, zoom into the upper hemisphere showing the
systematic deviation between velocity (yellow to red) and normal vectors
(yellow). Right: Simulation results for $\alpha$, the angle with
the velocity direction, as a function of $\theta$, for different
$v_{0}$ (legend) and $J$. From steep to shallow: $J=0.1\tau^{-1}$, with
fitted $s=1.25$, $J=1\tau^{-1}$ with $s=0.45$ and $J=10\tau^{-1}$ with $s=0.15$.
\label{fig:alpha_theta} }
\end{figure}

Then we can already derive a simple relation between the rotation
speed and the active forces by projecting equation (\ref{eq:projection2})
onto the $\phi$ direction. Assuming a steady state rotation of our
slice, we have $\dot{\mathbf{r}}_{i}=R\omega\cos\theta_i \mathbf{e}_{\phi}$,
and if the $\phi$ projections of any interparticle forces on particle
$i$ cancel out, we are left with: 
\begin{equation}
\cos\theta_{i}=\frac{v_{0}}{R\omega}\cos\alpha_{i}.
\end{equation}
In this simplest case, the solution is radially symmetric. We can
easily solve for $\alpha_i$, \emph{if and only if} we assume that the
velocity at the equator equals the self propulsion speed $v_{0}$.
Then $\frac{v_{0}}{R\omega}=1$, and we have $\alpha_{i}=\theta_{i}$;
that is a pattern where $ \mathbf{n}_{i}$ is parallel to the
direction of motion along the equator, and pointing inwards symmetrically
on both sides, as can be seen in Figure \ref{fig:alpha_theta} (left).

\begin{figure*}[t]
\centering
\includegraphics[clip,width=0.9\textwidth]{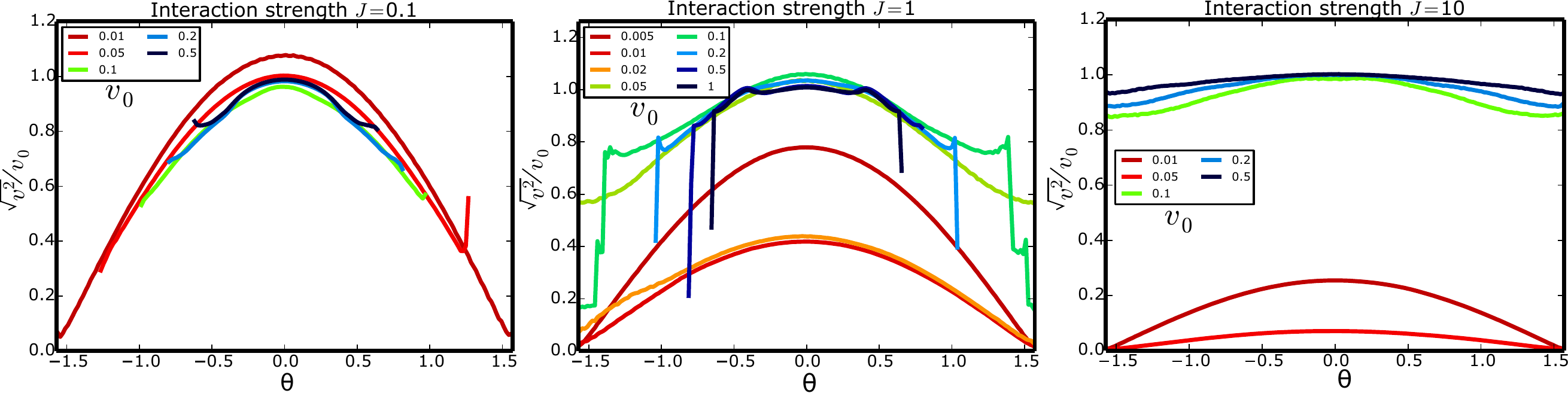}
\protect\caption{Velocity profiles for increasing $v_{0}$ and $J=0.1\tau^{-1}$ (left), $J=1\tau^{-1}$
(middle) and $J=10\tau^{-1}$ (right).\label{fig:velocity_profiles} }
\end{figure*}

In practice, we find that $\alpha=s\theta$, with a slope $s$ that
is nearly independent of $v_{0}$ and only depends on $J$, as shown
in Figure \ref{fig:alpha_theta}, right. We fit the three values of $J$ that we explored as follows: $J=0.1\tau^{-1}$, with
fitted $s=1.25$, $J=1\tau^{-1}$ with $s=0.45$ and $J=10\tau^{-1}$ with $s=0.15$. The velocity profiles themselves
are also relatively complex, see Figure \ref{fig:velocity_profiles}.
The parabolic profiles at low $v_{0}$ are consistent with perfect
block rotation at $\omega=v(\theta=0)/R$. For $J=0.1\tau^{-1}$, this rotation
speed is simply $v_{0}/R$, however it is lower at larger $J$, indicating
a new type complex slowing down dynamics which deserves to be explored.
Once the band develops, the profiles become more complex, but for
$J=0.1\tau^{-1}$ and $J=1\tau^{-1}$ they retain sufficiently close similarity to a
parabolic profile in the region where the density is nonzero for the
approximation $\omega=v_{0}/R$ to remain valid.

Along the chain now, in the direction $ \mathbf{e}_{\theta}$,
we can approximate the solution following a force-on-a-chain method.
Suppose that our orange slice has $N_{p}$ particles in total from
pole to pole. Assuming overlapping particles, the force an adjacent
particle $j$ exerts on particle $i$ in the chain is given by $\mathbf{F}_{ij}=-k\hat{\mathbf{r}}_{ij}(2\sigma-|\mathbf{r}_{j}-\mathbf{r}_{i}|)$.
Here $k$ is the (linearized) stiffness of the potential and $\sigma$
is the particle radius. We can introduce curvilinear coordinates along
the chain. Expressed using angles, we have $\mathbf{r}_{j}-\mathbf{r}_{i}=R\left(\cos\theta_{j}-\cos\theta_{i},0,\sin\theta_{j}-\sin\theta_{i}\right)$.
If we expand around $\theta_{i}$ in small values of $\delta\theta=\theta_{j}-\theta_{i}$,
$\mathbf{r}_{j}-\mathbf{r}_{i}=-R\left(\sin\theta_{i}\delta\theta,0,-\cos\theta_{i}\delta\theta\right)$,
or finally $\mathbf{r}_{j}-\mathbf{r}_{i}=-R(\theta_{j}-\theta_{i}) \mathbf{e}_{\theta}$.

To first order, interparticle forces are along $ \mathbf{e}_{\theta}$,
and the forces acting on particle $i$ from its neighbors $i-1$ and
$i+1$ are $F_{i,i-1}=k(2\sigma-R(\theta_{i}-\theta_{i-1}))$ and
$F_{i,i+1}=-k(2\sigma-R(\theta_{i+1}-\theta_{i}))$.

Finally we can then write down the equations of motion along the chain:
\begin{align}
  0=&-v_{0}\sin\alpha_{1}-\mu k\left(2\sigma-R(\theta_{2}-\theta_{1})\right)\nonumber \\
  0=&-v_{0}\sin\alpha_{i}+\mu k\left(2\sigma-R(\theta_{i}-\theta_{i-1})\right) \nonumber \\
 &-\mu k\left(2\sigma-R(\theta_{i+1}-\theta_{i})\right) \label{eq:band_balance_supp} \\
 0=&-v_{0}\sin\alpha_{N}+\mu k\left(2\sigma-R(\theta_{N}-\theta_{N-1})\right)\nonumber.
\end{align}

\section{Approaches to a solution of equation (\ref{eq:band_balance_supp})}

\subsection{Continuum model: Eulerian vs. Lagrangian pictures}

If our system is large, i.e. if $\kappa=2\sigma/R\ll1$, where $\kappa$
is the dimensionless curvature of the sphere, the angular differences
can be written in differential form. Let the $u_{i}$ be the deviations
of the chain particles from their rest state, i.e. $\theta_{i}=\theta_{i}^{0}+u_{i}$,
with $\theta_{i}^{0}=\frac{2\sigma i}{R}-\frac{\sigma N}{R}$.

This transformation needs to be done carefully, and we can use either
an absolute reference frame or the coordinates of the particles themselves.
Let $\vartheta$ be the underlying angular coordinate we would like
to use for our solutions, with $\vartheta=0$ at the equator. Since
we use an absolute coordinate system, and not the particles themselves
for coordinates, this approach is in the Eulerian picture (Chaikin
and Lubensky, p.330-331)  \cite{chaikin2000principles}. Conversely,
if we use the original positions of the particles, $\theta_{0}$,
as a reference, the approach is Lagrangian. Habitually, Eulerian and
Lagrangian elasticity are defined as follows. Let $R$ be the original
positions in the undistorted material. Then after distortion, their
coordinates are given by $x(R)=R+u(R)$, where the initial positions
$R$ as used as reference frame. Lagrangian elasticity is based on
on this approach: distances in the distorted material are expressed
as $dx^{2}-dR^{2}=2u_{ij}^{L}(R)dR_{i}dR_{j}$, where $u_{ij}^{L}(R)$
is the Lagrangian strain tensor, 
\begin{equation}
u_{ij}^{L}(R)=\frac{1}{2}\left[\frac{\partial u_{i}}{\partial R_{j}}+\frac{\partial u_{j}}{\partial R_{i}}+\frac{\partial u_{k}}{\partial R_{i}}\frac{\partial u_{k}}{\partial R_{j}}\right].\label{eq:Lagrangian_strain}
\end{equation}
In an Eulerian approach, we use the new coordinates $x$ in the absolute
reference frame as a basis, and we have to invert the relation above
to have $R(x)=x-u(R(x))$ which then leads to the Eulerian strain
tensor $dx^{2}-dR^{2}=2u_{ij}^{E}(x)dx_{i}dx_{j}$. The Eulerian strain
tensor has a minus sign in the nonlinear term, opposite to the more
familiar Lagrangian strain tensor: 
\begin{equation}
u_{ij}^{E}(x)=\frac{1}{2}\left[\frac{\partial u_{i}}{\partial x_{j}}+\frac{\partial u_{j}}{\partial x_{i}}-\frac{\partial u_{k}}{\partial x_{i}}\frac{\partial u_{k}}{\partial x_{j}}\right].\label{eq:Eulerian_strain}
\end{equation}
For us, the initial relation $x(R)=R+u(R)$ is simply $\vartheta=\theta_{i}=\theta_{i,0}+u_{i}$,
which we then need to invert to obtain $R(x)=x-u(R(x))$, i.e. $\theta_{i,0}(\vartheta)=\vartheta-u(\theta_{i,0}(\vartheta))$.
Our strain tensor is affected by the one-dimensional nature of our
problem. By definition, the metric tensor has to be a perfect square
for a one dimensional problem, $dx^{2}=g^{L}(R)dR^{2}$, so that $g^{L}(R)=(1+du/dR)^{2}$,
and $dR^{2}=g^{E}(x)dx^{2}$ with $g^{E}(x)=(1-du/dx)^{2}$. In our
coordinates, we then derive the strain tensors: 
\begin{align}
 & u_{s}^{L}=\frac{du}{d\theta^{0}}+\frac{1}{2}\left(\frac{du}{d\theta_{0}}\right)^{2}\\
 & u_{s}^{E}=\frac{du}{d\vartheta}-\frac{1}{2}\left(\frac{du}{d\vartheta}\right)^{2}.
\end{align}

To recover the underlying periodicity (the $i$ index), recall the
standard definition of a reciprocal vector $G$ for a lattice: $G\cdot R=2\pi m$,
with $m$ and integer. For us $G\cdot\theta_{i,0}=2\pi i$. In the
$\vartheta$ basis, the old positions of the undistorted lattice points
still have to follow $G\cdot(x-u(x))=2\pi m$, that is for us then
$G\cdot(\vartheta-u(\vartheta))=2\pi i$; or using the lattice definition
of the $\theta_{i,0}$, $\theta_{i,0}(\vartheta)=\vartheta-u(\vartheta)=\frac{2\sigma i}{R}-\frac{\sigma N}{R}$.

For the Lagrangian coordinates, the transformation to continuum is then straightforward:
We can approximate the angle differences as $\theta_{i}-\theta_{i-1}=\theta_{i,0}-\theta_{i-1,0}+u(\theta_{i,0})-u(\theta_{i-1,0})\approx\frac{2\sigma}{R}+\frac{2\sigma}{R}\frac{du}{d\theta_{0}}$.
The double angle difference $\theta_{i+1}+\theta_{i-1}-2\theta_{i}=\theta_{i+1,0}-\theta_{i-1,0}-2\theta_{i,0}+u(\theta_{i+1,0})+u(\theta_{i-1,0})-2u(\theta_{i,0})\approx\left[\frac{2\sigma}{R}\right]^{2}\frac{d^{2}u}{d\theta_{0}^{2}}$
becomes now clearly a discrete Laplacian. In Eulerian coordinates,
the complexity arises from the difference in line element inherent
in passing to the new coordinates $\vartheta$. Though we clearly have
above $d\theta_0=\theta_{i,0}-\theta_{i-1,0}=2\sigma/R$, in the new coordinates we need to express it as a function of the new line element $d\vartheta$,
$d\theta_{0}=\sqrt{g^{E}}d\vartheta$, or more explicitly $d\vartheta=\left(1+\frac{du}{d\vartheta}\right)d\theta_{0}(\vartheta)$.
Then the angle differences become $\theta_{i}-\theta_{i-1}\approx\frac{2\sigma}{R}+\frac{2\sigma}{R}\left(1+\frac{du}{d\vartheta}\right)\frac{du}{d\vartheta}$
and $\theta_{i+1}+\theta_{i-1}-2\theta_{i}\approx\left[\frac{2\sigma}{R}\right]^{2}\left(1+\frac{du}{d\vartheta}\right)^{2}\frac{d^{2}u}{d\vartheta^{2}}$,
a much more complex set of derivatives.

Finally, the influence of the active Coriolis force still acts at
the \emph{distorted} points $\theta_{i}$. We can formally write $\theta_{i}=\theta_{i,0}(\vartheta)+u(\theta_{i,0}(\vartheta))$
in Eulerian coordinates, to see just as quickly that we just get $\theta_{i}=\vartheta-u(\theta_{i,0}(\vartheta))+u(\theta_{i,0}(\vartheta))=\vartheta$;
simply the angular coordinate. This makes sense since the active Coriolis
force is solely due to the constrained motion in the curved reference
frame, and completely independent of the initial particle positions.
In the Lagrangian frame, we need to keep track of the displacements
from the origin: $\theta_{i}=\theta_{i,0}+u(\theta_{i,0})$. The active
force contribution becomes tractable if we use the results from Figure
\ref{fig:alpha_theta} and assume that $\alpha=s\vartheta$, or equivalently
$\alpha=s\theta_{0}+su$. In the Lagrangian reference frame, we have
the equations

\begin{equation}
\frac{d^{2}u}{d\theta_{0}^{2}}=\alpha\sin\left(s\theta_{0}+su\right),\label{eq:Lagrange_bulk}
\end{equation}
with boundary conditions 
\begin{align}
 & \left.\frac{du}{d\theta_{0}}\right|_{-\theta_{m,0}}=\beta\sin\left(s\theta_{0}+su\right)\nonumber \\
 & \left.\frac{du}{d\theta_0}\right|_{\theta_{m,0}}=-\beta\sin\left(s\theta_{0}+su\right).
\end{align}

In the Eulerian reference frame, the right hand side term is simpler;
however additional derivatives arise on the left hand side: 
\begin{equation}
\left(1+\frac{du}{d\vartheta}\right)^{2}\frac{d^{2}u}{d\vartheta^{2}}=\alpha\sin(s\vartheta),\label{eq:Euler_bulk}
\end{equation}
with boundary conditions 
\begin{align}
 & \left.\left(1+\frac{du}{d\vartheta}\right)\frac{du}{d\vartheta}\right|_{-\vartheta_{m}}=\beta\sin(s\vartheta)\nonumber \\
 & \left.\left(1+\frac{du}{d\vartheta}\right)\frac{du}{d\vartheta}\right|_{\vartheta_{m}}=-\beta\sin(s\vartheta).
\end{align}
Here, $\alpha=\frac{1}{R}\left[\frac{R}{2\sigma}\right]^{2}\frac{v_{0}}{\mu k}$,
and $\beta=\frac{v_{0}}{2\sigma\mu k}$, and the boundary conditions
have to be taken at the original position of the chain edges $\theta_{m,0}$
in the Lagrangian case, but at the final position $\vartheta_{m}$
for the Eulerian equations. The two approaches are strictly equivalent,
as can be seen by applying a change of variable $\theta_{0}=\vartheta-u$
and $d\theta_{0}=\left(1+\frac{du}{d\vartheta}\right)d\vartheta$
to the Lagrangian equations.

\begin{figure}[h]
\includegraphics[clip,width=0.99\columnwidth]{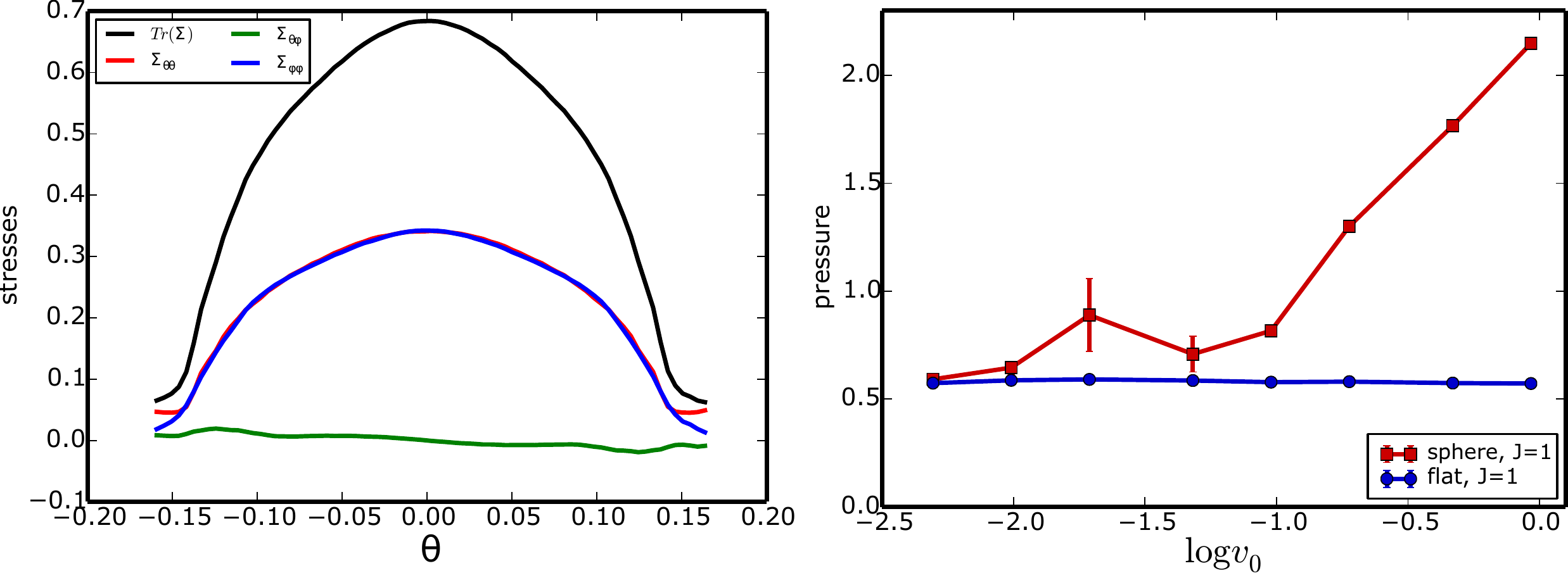}
\protect\caption{Left: Averaged profiles of the components of the stress or force moment
tensor in the local frame $\mathbf{e}_r$, $\mathbf{e}_{\theta}$, $\mathbf{e}_{\phi}$,
for $J=1\tau^{-1}$ and $v_{0}=1\sigma/\tau$. The stress tensor is close to isotropic
on the sphere, with $\Sigma_{\theta\theta}\approx\Sigma_{\phi\phi}$,
and the off-diagonal components are much smaller. Due to the projection,
all of the components involving $\mathbf{e}_r$ are zero. Right:
Mean pressure for the spherical and the flat case as a function of
$v_{0}$. The active part of the pressure is only significant compared
to the overlap part of the pressure $p_{0}\approx0.5$ for the spherical
case; in other words activity does not induce energy-storing distortions
in the flat case.}
\label{fig:stress_analysis} 
\end{figure}

\subsection{Predicting strain, pressure and density profiles}

Neither of the two systems of equations, unfortunately, has an analytical
solution. From here on we proceed with approximations. Below, we compare
the approximate results to a discrete energy minimization approach
for the chain, and show that they are valid in the low $v_{0}$ regime.
In both the Lagrangian and Eulerian case, the $0^{th}$ order equation
that can be solved is $\frac{d^{2}u}{d\theta^{2}}=\alpha\sin(s\theta)$,
with boundary condition $\frac{du}{d\theta}|_{\theta_{m}}=-\beta\sin(s\theta_{m})$
and equivalently at $-\theta_{m}$. The solution to this equation
is $u(\theta)=-\alpha/s^{2}\sin s\theta+c\theta+d$, where $c$
and $d$ are integration constants. We can immediately see that the
symmetries $u(-\theta)=-u(\theta)$ and $u(0)=0$ require that $d=0$.
Clearly, the two boundary conditions are equivalent, and we are left
with $-\alpha/s\cos s\theta_{m}+c=-\beta\sin s\theta_{m}$ to determine
$c$.

We still lack a relation tying $\theta_{m}$ to the underlying physics
of the chain. In the Lagrangian case, $\theta_{m,0}$ is simply the
initial extent of the chain before the active forces are applied.
Since at our high density, the sphere is covered in particles in the
absence of driving, we can safely assume $\theta_{m,0}=\pi/2$. In the Eulerian
case, this is slightly more tricky. Consider the elementary differential
geometry relation for a curve $\mathcal{C}$ parametrized by $\mathbf{l}(t)$
in space $S$. Its length is given
by $L=\int_{\mathcal{C}}\sqrt{\sum_{k}(dl_{k}/dt)^{2}}dt$ (e.g.,
p.95 Jean Schmets, `Introduction au calcul integral')  \cite{Schmets1994analysis}.
In our Eulerian approach, $\vartheta=t$, the parametrization, and
the mapped space $S$ belongs to the original $\theta_{0}=\mathbf{l}$, and where
$L=\pi $ is the original length of the chain. If this seems backwards,
it is compared to a more standard Lagrangian parametrization, where it would be the
other way round. The set of derivatives are now simply the square
root of the metric tensor, $d\theta_{0}/d\vartheta=(1-du/d\vartheta)=\sqrt{g^{E}(\vartheta)}$.
Then the missing equation linking the original chain length and the
displacement field is: 
\begin{equation}
\pi=\int_{-\vartheta_{m}}^{\vartheta_{m}}d\vartheta\left(1-\frac{du}{d\vartheta}\right)
\end{equation}

This last equation does not have an analytical solution, and the the
approximate solution to the chain profile in Eulerian coordinates
can only be given implicitly: 
\begin{align}
& u(\vartheta)=  -\alpha/s^{2}\sin s\vartheta+(\alpha/s\cos s\vartheta_{m}-\beta\sin s\vartheta_{m})\vartheta\label{eq:displacements_Euler}\\
& \pi= \frac{2\alpha}{s^{2}}\sin(s\vartheta_{m})+2\theta_{m}\left[1-\alpha/s\cos(s\vartheta_{m})+\beta\sin(s\vartheta_{m})\right]\label{eq:thetam_implicit}\\
& \alpha= \frac{1}{R}\left[\frac{R}{2\sigma}\right]^{2}\frac{v_{0}}{\mu k},\quad\beta=\frac{v_{0}}{2\sigma\mu k},\label{eq:constants}
\end{align}
In the Lagrangian case, at the $0th$ level the solution is simpler:
\begin{equation}
u(\theta_{0})=-\alpha/s^{2}\sin s\theta_{0}+\left[\alpha/s\cos\frac{s\pi}{2}-\beta\sin\frac{s\pi}{2}\right]\theta_{0},
\end{equation}
however, to compare to simulation results, all expressions have to
be evaluated at the new positions $\vartheta=\theta_{0}+u(\theta_{0})$.

The Eulerian strain is given by 
\begin{align}
u_{s}^{E}(\vartheta) & =\frac{du}{d\vartheta}-\frac{1}{2}\frac{du}{d\vartheta}\frac{du}{d\vartheta}\nonumber \\
 & \approx\!-\!\frac{1}{2R}\!\!\left[\frac{R}{\sigma}\right]^{2}\!\!\frac{v_{0}}{\mu k}\left[\frac{1}{s}(\cos s\vartheta\!-\!\cos s\vartheta_{m})\!+\!\frac{2\sigma}{R}\sin s\vartheta_{m}\!\right]\!,\label{eq:strain_profile}
\end{align}
where we have only kept the first order strain term $\frac{du}{d\vartheta}$
in the second equation. This is equation (4) in the main text. In Lagrangian coordinates, at the first order,
we have the exact same expression, except using $\theta_{0}$ instead
of $\vartheta$ and $\pi/2$ instead of $\theta_{m}$. 

We can estimate the pressure profile within the dense phase by noting
that the interparticle forces are related to the derivative of the
displacement profile: 
\begin{equation}
F_{i,i+1}=-k\left(2\sigma-R(\theta_{i+1}-\theta_{i})\right)=2k\sigma\frac{du}{d\theta}.
\end{equation}
This is assuming that all the $F_{i,i+1}=0$ before any displacements
were applied; or in other words we have no pre-stress in the system.
We discuss the evidence for pre-stress and its implications in the
next section.

The interaction part of the stress tensor at the local scale is given
by 
\begin{equation}
\hat{\sigma}_{i}=\frac{1}{A_{i}}\sum_{j}\mathbf{r}_{ij}\mathbf{F}_{ij},\label{eq:microstress}
\end{equation}
where the $\mathbf{r}_{ij}$ reach from the centre of each particle
to the contact and $A_{i}$ is the part of an area tessellation (e.g.
Voronoi diagrams) belonging to particle $i$  \cite{ball_stress_2002,irving_statistical_1950}.
Ignoring second order contributions in $u$, we estimate $r_{ij}\approx\sigma$
and $A_{i}\approx4\sigma^{2}$. If each particle has four contacts,
and horizontal forces equal vertical forces (i.e. the stress field
is isotropic), the pressure is given by $p_{i}=\text{Tr}\hat{\mathbf{\sigma}}_{i}=2k\frac{du}{d\vartheta}$
(note the units of force / length, or stiffness, appropriate to two
dimensions). This is really just a microscopic derivation of the stress-strain relation; and we should write $p=\bar{k}u_{s}(\vartheta)$,
with a possibly effective stiffness constant $\bar{k}$. 

To test our
assumption of an isotropic stress field, we analysed the components
of the force moment tensor in the local frame $\mathbf{e}_r$, $\mathbf{e}_\theta$, $\mathbf{e}_{\phi}$
($\hat{\Sigma}_{i}=A_{i}\hat{\sigma}_{i}$ is the additive version of the stress
tensor, with units of energy). For an isotropic stress tensor $\Sigma_{\theta\theta}\approx\Sigma_{\phi\phi}$,
the off-diagonal components are much smaller and due to the projection,
all of the components involving $\mathbf{e}_r$ are zero. Figure
\ref{fig:stress_analysis} (left) shows that in a developed band, these approximations
hold to a very high degree.

\begin{figure*}
\includegraphics[clip,width=0.9\textwidth]{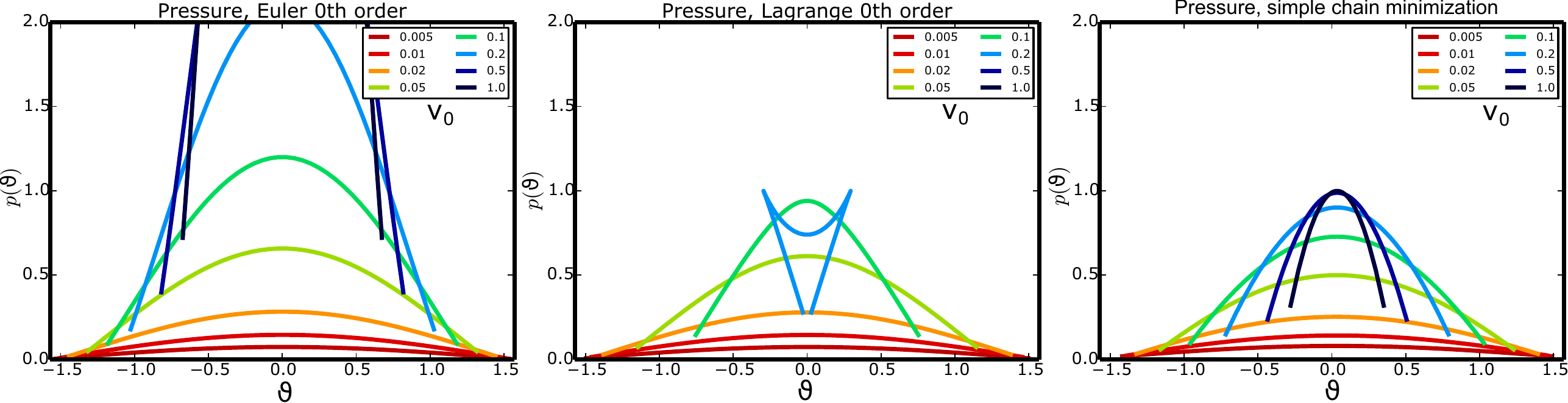}
\protect\caption{Comparing the analytical solution to energy minimization. Left: Analytical
$0^{th}$ order Eulerian solution using the full Eulerian strain tensor.
Middle: Analytical $0^{th}$ order Lagrangian solution, starting from
the full Lagrangian strain tensor. Right: Predicted pressure of the
single-overlap chain with $N=44$ (just touching) particles. Note
the good agreement at low values of $v_{0}$. The remaining parameters
are $s=1$, $p_{0}=0$, $R$ and $\sigma$ as in the simulation.}

\label{fig:analytics} 
\end{figure*}

Then the predicted pressure profile for $|\vartheta|<\vartheta_{m}$ is
(again, neglecting the second order contributions to the strain):
\begin{equation}
p(\vartheta)=-\frac{v_{0}R}{2\mu\sigma^{2}}\left[\frac{1}{s}(\cos s\vartheta+\cos s\vartheta_{m})-\frac{2\sigma}{R}\sin s\vartheta_{m}\right],\label{eq:pressure_profile}
\end{equation}
with an equivalent expression for the Lagrangian pressure profile.
A couple of interesting remarks: 
\begin{itemize}
\item The pressure is negative for $|\vartheta|<\vartheta_{m}$, i.e., this
is a compressive stress (we plot $-p$ throughout). 
\item The pressure is not zero when the edge of the hole is reached at $\vartheta_{m}$.
This is due to the contribution of the active driving forces which
lead to an inwards pressure of $2v_{0}\sin\vartheta_{m}/\mu\sigma$.
This is consistent with the boundary conditions, and comes from the
$\sum_i\mathbf{r}_{i}\mathbf{F}_{i}/A_i$ non-pair forces part of the
Irving-Kirkwood stress tensor  \cite{irving_statistical_1950}. In fact, this is identical with the estimate of the
active pressure that the gas phase exerts on the cluster phase in
studies of the first order clustering transition of self-propelled
particles~  \cite{fily2013freezing}. 
\item The pressure does not depend on $k$; that is it becomes independent
of the details of the interaction, and depends instead only on the
dynamical parameter $v_{0}/\mu$ and the geometrical parameters $\sigma$
and $R$. 
\end{itemize}
Finally, we can also predict the angular density profile: We define
the local density to be $\rho=1$ when particles are just touching
(i.e. the unperturbed chain). Then, assuming again isotropic compression
like for the pressure profile above, $\rho\approx1+|\frac{du}{d\bar{\theta}}|$,
or more precisely using the strain $\rho(\vartheta)=1-u_{s}(\vartheta)$.
The density profile is $\rho=0$ for $|\vartheta|>\vartheta_{m}$
and to first order we have
\begin{equation}
\rho(\vartheta)\!=\!1-\frac{1}{2R}\!\!\left[\frac{R}{2\sigma}\right]^{2}\!\!\!\frac{v_{0}}{\mu k}\!\left[\frac{1}{s}(\cos s\vartheta\!-\!\cos s\vartheta_{m})+\frac{2\sigma}{R}\sin s\vartheta_{m}\right]\!,
\end{equation}
otherwise. Interestingly, unlike the pressure, the density depends
on $k$ and doesn't seem to be universal. Again, there is a similar
equivalent equation for the Lagrangian density prediction.

Figure \ref{fig:analytics} (left and middle) show the analytical
predictions for the pressure profiles (plotting $-p$) using the \emph{full} Eulerian
and Lagrangian strain tensors, evaluated at the simulation parameters
for $R$, $\sigma$ and $v_{0}$ and using Mathematica to numerically solve the
implicit equation for $\vartheta_{m}$. We have also used $s=1$,
and ignored any pre-stress contributions. While the profiles agree
with each other at low $v_{0}$, there are considerable differences
at higher $v_{0}$; the Lagrangian solution also stops being single-valued
due to $u\gg\theta_{0}$ in evaluating $\vartheta$.

\section{Discrete chain models based on energy minimization}

Given the large discrepancy between the Eulerian and Lagrangian approximate
analytical solutions, it becomes clear that a numerical approach is
inevitable. Instead of numerically solving the full equations, which
cannot incorporate the full effects of discreteness, we use an energy
minimization type of approach. We can treat equations (\ref{eq:band_balance_supp})
as Euler-Lagrange equations of an energy functional containing only
potential energy terms. Formally, even though our physical system
conserves neither energy nor momentum, if we assume $\alpha=s\theta$,
the active force components in equation (\ref{eq:band_balance_supp}) derive
from an effective potential $V_{\text{act}}^{i}=v_{0}\cos(s\theta_{i})$
which can be added to the interparticle repulsive term $V_{\text{rep}}^{i}=\frac{kR}{2}\sum_{j\in\mathcal{N}}(\theta_{j}-\theta_{i})^{2}$,
where we initially only consider nearest neighbors. Then setting the
gradients of $V^{i}=V_{\text{act}}^{i}+V_{\text{rep}}^{i}$ to zero
is equivalent to equations (\ref{eq:band_balance_supp}). We then minimize
the potential by using the standard L-BFGS-B conjugate gradient method,
and compute strain and pressure from the numerically evaluated displacements
via the route discussed above. In Figure \ref{fig:analytics}, left,
we show the resulting pressure profiles for the same set of parameters
as the analytical results in the two other plots. Analytics and energy
minimization agree with each other in the region $v_{0}\ll1$
where the approximation of small displacements remain valid.

\section{Numerical comparison to simulation}

In our simulation, we keep most system parameters fixed, and instead
vary the dynamical parameters $v_{0}$ and $J$. The alignment parameter
$J$ only appears through its influence on the parameter $s$, with
$s$ reducing for larger alignment strengths. Then the main remaining
dynamical parameter is simply $v_{0}$. Our constant parameters are
$\mu=1$, $\phi=1$, $\sigma=1$ and $R=28.2094791\sigma$, or equivalently
$N_{\text{tot}}=3183$ and the stiffness constant $k=1$. An important
parameter is the dimensionless curvature $\kappa=2\sigma/R=0.0708982$,
which will be our small parameter in expansions (note that $\alpha=\beta/\kappa$).
From our analysis of the $\alpha$-$\theta$ relation, we retain the
fit values $s=1.25$ for $J=0.1\tau^{-1}$, $s=0.45$ for $J=1\tau^{-1}$ and finally
$s=0.15$ for $J=10\tau^{-1}$. Finally, we estimate our main parameters as:
\begin{align}
\alpha & =\frac{1}{R\kappa^{2}}\frac{v_{0}}{\mu k}=7.05237v_{0}\\
\beta & =\frac{1}{2\sigma}\frac{v_{0}}{\mu k}=0.5v_{0}.\label{eq:parameters}
\end{align}

An important issue is to determine the correct initial state for the
chain. Since our packing fraction $\phi=1$, one might think that
just touching spheres with no pre-stress are the correct initial state.
However, $\phi=1$ is in the jammed or crystalline region of phase
space, where soft particles interpenetrate, and our sphere is no exception.
In Figure \ref{fig:stress_analysis} (right), we show the mean pressure
(or to be precise, the trace of the force-moment tensor) in the spherical
system as a function of $v_{0}$, and compare it to the pressure in
an equivalent flat system. For the flat system, the pressure is very
close to constant, indicating no strain-inducing distortions due to
activity, consistent with the observed block-translation in these
cases. The constant value $p_{0}=0.5k/\sigma^2$ stems purely from the overlaps
of the particles due to the initial packing. If we assume $\bar{z}=6$
neighbors on average, we can estimate an initial overlap of roughly
$\delta_{0}=0.1\sigma$. To make a quantitative comparison between the chain
model and the simulation, we need the same starting value of $p_{0}$
and so we prepare the chain with initial overlaps of $\delta=0.25\sigma$. This
is equivalent of a chain length of $N_{p}=59$. The actual number
of particles in a chain can be estimated by straightforward counting
in Figures 3 and 4{\bf a} of the main text and gives an
estimate of $N_{p}=38,35$ and $32$ for the steady states at $v_{0}=0.03\mu k \sigma$,
$0.1\mu k \sigma$ and $1\mu k \sigma$.

When comparing the simple chain minimization results such as Figure
\ref{fig:analytics} (right) to the simulation, it soon becomes apparent
that it dramatically underestimates the pressure in the centre of
the band. This points to a larger effective stiffness constant $\bar{k}>k$
in the centre. The most straightforward explanation for this is double
or even multiple overlaps of particles, \emph{i.e.}~next-nearest neighbor
and further interactions. We have confirmed their existence in the
simulated bands, and so incorporated them into the chain minimization
procedure by counting \emph{all} neighbors in the repulsive term $V_{\text{rep}}^{i}=\frac{kR}{2}\sum_{j\in\mathcal{N}}(\theta_{j}-\theta_{i})^{2}$.
Due to the initial compressed state, we also add a constraint $0\leq\theta_{i}\leq\pi$
to the L-BFGS-B minimization routine. Finally, with this amount of
detail, the continuum formulations $p=\bar{k}u_{s}$ and $\rho=1+u_{s}$
lose their meaning and we directly compute the pressure via the force
moment tensor and the density through a histogram.

\begin{figure*}
\includegraphics[clip,width=0.99\textwidth]{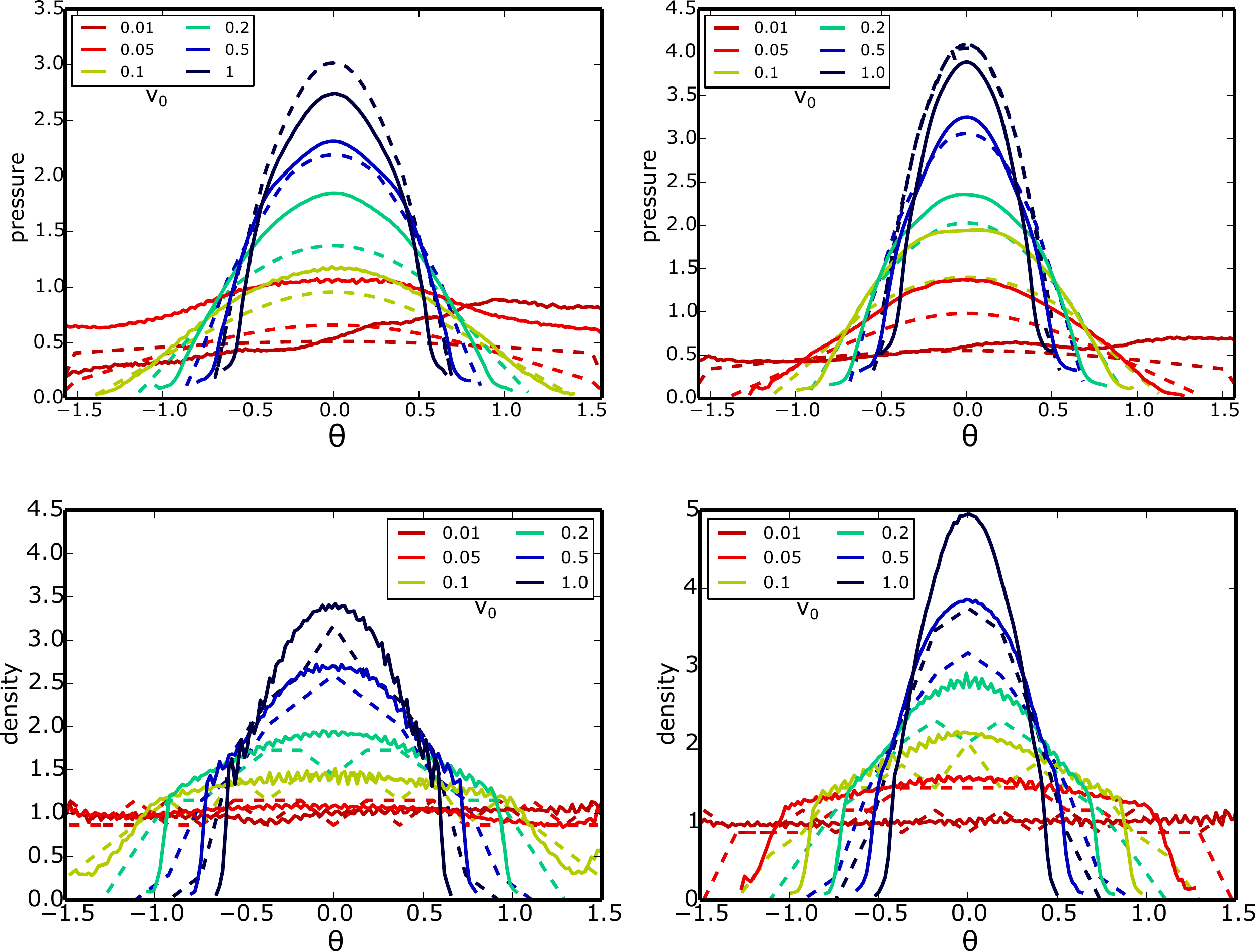}
\protect\caption{Simulation results (solid) and full chain calculation (dashed) compared
to each other, for $J=1\tau^{-1}$ and $s=0.6$ (left) and $J=0.1\tau^{-1}$ and $s=1.25$ (right). Top row: pressure, and bottom row: density}
\label{fig:chain_fit} 
\end{figure*}

Figure \ref{fig:chain_fit}
shows the numerical stress and density profiles for $J=0.1\tau^{-1}$ and $J=1\tau^{-1}$,
overlaid with the full chain minimization results. We have used $s=1.25$
for $J=0.1\tau^{-1}$ as fitted, but had to adjust $s=0.6$ for $J=1\tau^{-1}$, indicating
that the chain model approximations work better for a narrow peaked
band. The model provides a good quantitative fit for both sets of
simulation. The numerical chain minimization results for $J=1\tau^{-1}$ are
shown as a standalone graph in Figure 4 (f) and (g) of the main paper,
and compare to the numerical profiles in Figure 4 (a) and (b).

\begin{figure*}
\includegraphics[clip,width=0.99\textwidth]{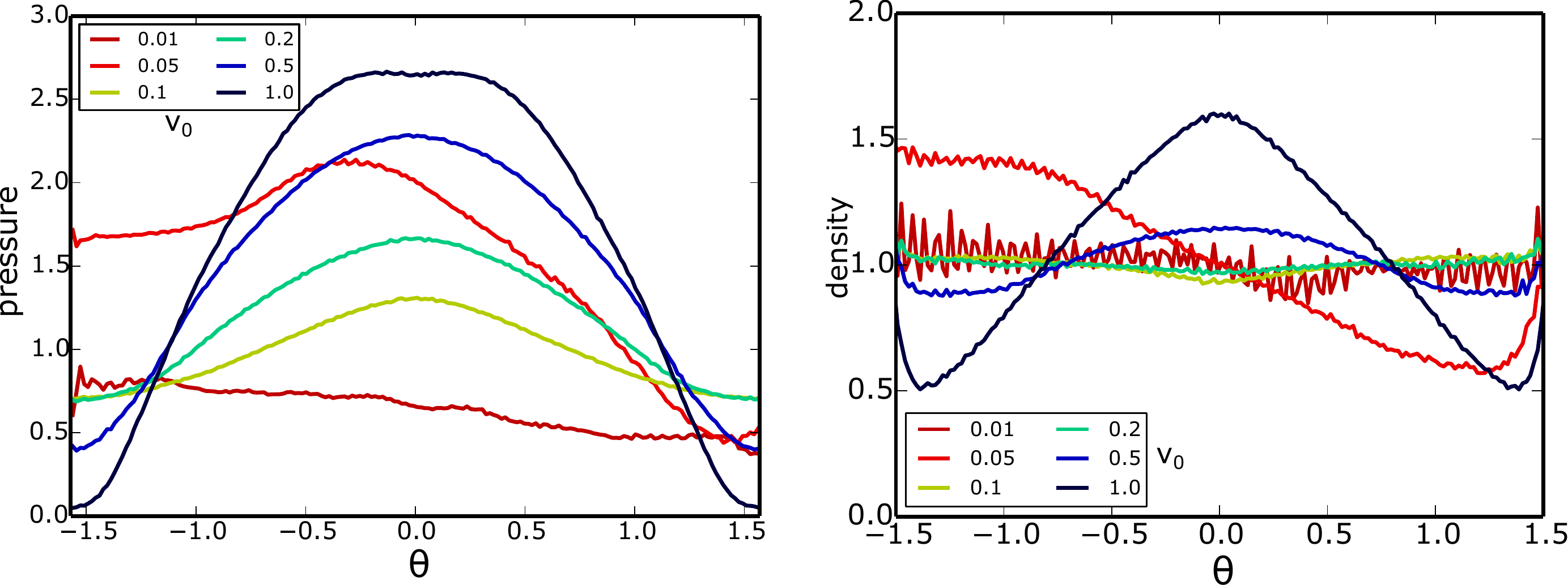}
\protect\caption{Density (left) and pressure (right) for $J=10\tau^{-1}$. In addition to the
unusually broad band developing at large $v_{0}$, a more complex
transition involving a unipolar symmetry seems to be taking place
at lower $v_{0}$. The full chain calculation is unable to reproduce
these pressure and density profiles.}
\label{fig:caseJ10} 
\end{figure*}

For $J=10\tau^{-1}$ (see Figure \ref{fig:caseJ10}), the peaked density profile
is slow to develop, and the pressure profile remains very broad at
all values of $v_{0}$. We were not able to satisfactorily fit them
with any value of $s$, especially not at $s=0.15$. We believe that
at low $v_{0}$ and large values of $J$, band order is slow to develop,
and we reach an intermediate phase where the recently discovered density
instability in repulsive self-propelled particles (see e.g.  \cite{fily2013freezing})
influences the behaviour. This conjecture is also supported by the
order parameter graph, Figure 3 (h) of the main publication, where
there is a consistent dip in the order parameter at intermediate values
of $v_{0}$ for the higher values of $J$.

% \bibliographystyle{unsrt}
% \bibliography{suppbiblio}

\end{document}